\documentclass[a4paper,fleqn]{cas-dc}

\usepackage[numbers]{natbib}

\usepackage{booktabs}
\usepackage{multirow}
\usepackage{graphicx}
\usepackage{amsmath}
\usepackage{enumitem}
\usepackage{booktabs}
\usepackage{subcaption}

\let\cite\citep

\def\tsc#1{\csdef{#1}{\textsc{\lowercase{#1}}\xspace}}
\tsc{WGM}
\tsc{QE}
\tsc{EP}
\tsc{PMS}
\tsc{BEC}
\tsc{DE}

\begin{document}
\let\WriteBookmarks\relax
\def\floatpagepagefraction{1}
\def\textpagefraction{.001}

\shorttitle{Energy Constrained Hierarchical Underwater Monitoring via Local Multi-Agent RAG}

\shortauthors{Mohamed Amine JANATI et~al.}

\title [mode = title]{Energy Constrained Hierarchical Underwater Monitoring via Local Multi-Agent RAG}
\tnotemark[1]

\tnotetext[1]{This document presents the results of the internship funded by the French Research Institute for Exploitation of the Sea (Ifremer).}

\author[1,2]{Mohamed Amine Janati}[
    bioid=1,
    orcid=0009-0002-5753-4567
]
\ead{mohamed-amine.janati@imt-atlantique.net}
\credit{Writing -- review \& editing, Writing -- original draft, Validation, Supervision, Conceptualization, Methodology, Visualization, Software, Data curation, Result analysis}

\author[1]{Laurent Gautier}[
    bioid=2,
    orcid=0000-0002-1501-6609
]
\cormark[1]
\credit{Writing -- review \& editing, Supervision, Conceptualization, Project administration, Funding acquisition, Validation, Resources}

\author[1]{Stéphane Barbot}[
    bioid=3,
    orcid=0000-0002-9051-6996
]
\credit{Writing -- review \& editing, Supervision, Conceptualization, Project administration, Funding acquisition, Validation, Resources}

\affiliation[1]{organization={Ifremer},
    addressline={1625 RTE de Sainte-Anne},
    city={Plouzané},
    postcode={29280},
    country={France}}

\affiliation[2]{organization={IMT Atlantique},
    addressline={Technopôle Brest-Iroise, CS 83818},
    city={Plouzané},
    postcode={29280},
    country={France}}

\cortext[1]{Corresponding author}

\begin{abstract}
Marine life monitoring is limited by strict energy constraints, poor underwater connectivity, and the high cost of transmitting raw multimodal data from remote deployments. This paper proposes a low-consumption underwater monitoring architecture that combines always-on edge sensing with selective high-performance local reasoning. The system follows a hierarchical master--satellite design in which ultra-low-power MAX78000/MAX78002 microcontrollers continuously monitor visual and acoustic signals, while an NVIDIA Jetson Orin NX is activated only for scheduled processing, event-driven analysis, or researcher interaction. Once active, the Jetson executes a fully local multimodal pipeline for data ingestion, visual target extraction, embedding-based indexing, species identification, retrieval-augmented reasoning, and automated reporting. BioCLIP/OpenCLIP embeddings are used to organize mission data, marine taxonomic references, scientific documents, and operational metadata in local ChromaDB collections. A dedicated identification layer combines visual similarity search, centroid-based classification, and supervised classifiers to support adaptive species recognition. A LangChain-based multi-agent framework coordinates query routing, structured analysis, energy management, hardware reconfiguration, and report generation. The architecture is evaluated through visual and acoustic monitoring case studies. The proposed system bridges ultra-low-power continuous sensing with local multimodal intelligence, enabling underwater stations to produce structured, researcher-ready knowledge while compressing local data for flexible acoustic, optical, or satellite transmission—minimizing both energy use and communication overhead.

\end{abstract}





\maketitle

\section{Introduction}
\label{sec:introduction}

Marine ecosystems are increasingly monitored using fixed and mobile observation platforms equipped with cameras, hydrophones sensors, and environmental probes. These systems generate large volumes of heterogeneous data that are essential for studying biodiversity, detecting ecological changes, and identifying unusual biological or geophysical events. However, underwater deployments remain constrained by limited energy availability, restricted communication bandwidth, difficult physical access, and the high cost of transferring raw multimodal data from remote sites. As a result, many monitoring workflows still rely on storing data locally and processing them only after recovery or delayed transmission, which limits real-time awareness, adaptive sensing, and rapid scientific interpretation.

Recent advances in embedded artificial intelligence offer new opportunities for local underwater data processing. Lightweight neural networks can be deployed on low-power microcontrollers to detect acoustic or visual events directly at the sensor level. At the same time, more capable edge processors can execute multimodal models, retrieval systems, and local language models for higher-level interpretation. Nevertheless, a major architectural challenge remains: continuously operating high-performance processors is often incompatible with long-duration underwater deployments, while ultra-low-power detectors alone cannot provide rich scientific reasoning, taxonomic identification, or contextual reporting. A practical monitoring station therefore requires a hierarchical design that combines continuous low-power sensing with selective activation of high-capacity local intelligence.

This paper proposes a low-consumption, multimodal, and agentic architecture for underwater monitoring, destined to operate autonomously on battery for severals months. The system follows a master--satellite topology in which two ultra-low-power microcontrollers act as always-on sentinel nodes, while an NVIDIA Jetson Orin NX acts as the high-performance master node. One MAX78002 is dedicated to visual event detection, while another MAX78000 microcontroller is used for acoustic monitoring. The sentinel layer performs continuous first-stage detection, whereas the Jetson is activated only when scheduled processing, event-driven analysis, or researcher interaction is required. This design reduces unnecessary energy consumption while preserving the ability to perform deeper multimodal analysis at the edge.

The proposed system supports three operating modes. In Autonomous Mode, the Jetson remains powered off and wakes periodically, for example once per day, to ingest and process detections accumulated by the sentinel nodes. In Hybrid Mode, relevant events detected by the visual or acoustic sentinels trigger immediate Jetson activation through a hardware wake-up interface, enabling lower-latency analysis. In Mission Mode, the Jetson remains available for direct interaction with researchers through a local interface, allowing queries over recent detections, indexed mission data, and taxonomic references. These modes allow the same architecture to support long-term autonomous deployments, event-driven monitoring, and field-based scientific inspection.

Once activated, the Jetson executes a fully local multimodal processing pipeline. Collected data are ingested, normalized, and indexed into local vector database. Images, video frames, scientific documents are indexed using an embedding model, in our case OpenCLIP or BioCLIP-2. For visual data, object localization can be performed using Grounding DINO or Yolo World to extract candidate marine targets before embedding. A dedicated species identification layer then operates in the embedding space using taxonomic centroids and supervised classifiers, such as Support Vector Machines, constructed from taxonomic reference datasets. This layer converts visual observations into taxonomic hypotheses that can be used for retrieval-augmented reasoning.

Beyond classical retrieval-augmented generation, the system includes a local multi-agent control framework. A RouterAgent selects the appropriate collection, modality, and retrieval pathway for each user query or system trigger. An AnalystAgent handles structured questions over video metadata using sandboxed Pandas-based computation, avoiding unnecessary language-model generation when deterministic analysis is sufficient. An Energy and Sensor Management Agent adapts sensing and compute behavior according to mission context and power constraints. A Hardware Configuration Agent selects specialized model binaries for the microcontroller sentinels, while a Reporting Agent produces compact Markdown summaries of indexed assets, detections, anomalies, energy state, and model status. These reports can be stored locally and transmitted later when connectivity becomes available.

The main contributions of this paper are as follows:
\begin{itemize}
\item \textbf{A hierarchical energy-aware underwater monitoring architecture} that combines MAX78000 \& MAX78002 low-power sentinel nodes with a Jetson Orin NX master node for selective high-performance processing.

\item \textbf{A fully local multimodal retrieval and reasoning pipeline} that indexes mission data, images, videos, scientific documents, taxonomic references, and operational metadata using BioCLIP-2/OpenCLIP embeddings and local ChromaDB collections.

\item \textbf{A dedicated species identification layer} that supports centroid-based taxonomic classification and supervised classification in the multimodal embedding space.

\item \textbf{A multi-agent control framework} that coordinates query routing, structured analytics, energy management, hardware reconfiguration, and autonomous scientific reporting directly at the edge.

\item \textbf{Integrated visual and acoustic monitoring workflows} implemented on ultra-low-power microcontrollers, including visual fish detection and marine mammal acoustic classification.

\item \textbf{In-situ data compression workflow} transforming raw underwater detections into compact embeddings for low-power transmission via acoustic, optical, or satellite networks.

\end{itemize}

By combining ultra-low-power continuous sensing with selective local multimodal reasoning, the proposed architecture enables underwater monitoring stations to reduce energy consumption and communication requirements while still producing structured, researcher-ready scientific information at the edge.

\section{Related Work}
\label{sec:related_work}

\subsection{Edge AI platforms from microcontrollers to embedded GPUs}
\label{subsec:edge_ai_platforms}

Edge AI for monitoring spans a wide computational spectrum, from milliwatt microcontrollers to embedded GPU platforms. At the lowest-power end, TinyML enables local inference under severe memory and energy constraints, making it suitable for always-on sensing nodes where communication or continuous high-resolution processing would dominate the energy budget \citep{Ray2022TinyML}. Hardware--software co-design frameworks such as MCUNet have shown that neural architecture search and optimized inference engines can make convolutional models feasible on microcontrollers with only hundreds of kilobytes of SRAM \citep{Lin2020MCUNet}. This direction is particularly relevant for autonomous marine monitoring, where long deployments require local preprocessing before storing or transmitting data.

Recent ultra-low-power object-detection work has further reduced the gap between microcontrollers and larger edge devices. TinyissimoYOLO targets object detection on highly constrained processors and was evaluated on platforms including the MAX78000, GAP9, STM32 and Apollo-class low-power microcontrollers \citep{Moosmann2023TinyissimoYOLOAICAS,Moosmann2024TinyissimoYOLOAccess}. Maxim / Analog Devices devices are especially relevant because the MAX78000 and MAX78002 integrate a convolutional neural-network accelerator while scoring very low on power consumption benchmarks\cite{millar2025benchmarkingultralowpowermunpus}. The MAX78002 extends this family with larger CNN memory and sensor-oriented interfaces such as MIPI CSI-2 and I2S, which are useful for compact camera and acoustic front-ends \citep{AnalogDevices2022MAX78002Datasheet}. 

Above the microcontroller tier, Raspberry Pi-class single-board computers provide a practical compromise between cost, programmability, and model complexity. They have been used in smart camera-trap and edge--cloud systems for real-time animal detection and species recognition, including continual-learning wildlife monitoring and ant-species detection with YOLO-based pipelines \citep{VelascoMontero2024CameraTraps,Palazzetti2025AntPi}. For more demanding underwater perception tasks, NVIDIA Jetson platforms provide GPU acceleration for real-time deep learning. For example, Jetson-based deep-sea crawler deployments have used YOLO models for automated species classification and counting in benthic imagery \citep{Ortenzi2024DeepSeaYOLO}. Another previous work carried by our team was the detection of invasive species, specifically LionFish, using Rasberry Pi module with a NPU accelerator \cite{laurent2025syrene}.These works show that each platform tier is useful, but they also motivate architectures that combine them: an always-on microcontroller can perform inexpensive event filtering, while a Raspberry Pi or Jetson can be activated only when richer visual, acoustic, or multimodal inference is required.

\subsection{Hierarchical cascade systems and event-triggered monitoring}
\label{subsec:cascade_monitoring}

A central challenge in autonomous monitoring is that the most accurate models are often too energy-intensive to run continuously. Hierarchical cascade systems address this problem by assigning low-cost decisions to low-power devices and escalating only uncertain or promising events to more capable processors. This principle appears in early wireless-sensor work on radio-triggered wake-up, where a low-power front-end activates a sleeping node only when communication or sensing is needed \citep{Gu2005RadioTriggeredWakeup}. Similar ideas have been applied to visual monitoring: sleepyCAM, for example, uses an external low-power trigger to wake or power a Raspberry Pi camera system, reducing the cost of continuous video surveillance \citep{Mekonnen2017SleepyCAM}. Such designs are directly compatible with edge AI cascades, where a microcontroller first runs a coarse detector or threshold model, then wakes a Raspberry Pi, Jetson, or communication module for expensive inference and reporting.

Marine monitoring deployments are strictly limited by battery, bandwidth, and storage, making tiered cascade processing essential for long-term operation. As demonstrated by real-time baleen-whale monitoring from ocean gliders \citep{Baumgartner2013Gliders}, a continuous, low-power first stage can screen raw acoustic or visual streams to flag candidate events, reserving resource-intensive high-tier processing, accurate multimodal detection, and data transmission strictly for detections of interest.

\subsection{Multimodal retrieval, vision--language models, and embedding-based species identification}
\label{subsec:multimodal_rag_species}

Vision--language models and contrastive embedding models provide a complementary direction for species identification. CLIP showed that image and text encoders trained with contrastive learning can support zero-shot recognition and retrieval by mapping images and natural-language labels into a shared embedding space \citep{Radford2021CLIP}. Biology-specific extensions make this paradigm more relevant for ecological applications. BioCLIP adapts vision foundation models to biological taxonomy and biodiversity imagery, while CLIBD extends contrastive learning toward the joint use of images, DNA barcodes, and taxonomic text for biodiversity monitoring \citep{Stevens2024BioCLIP,Gong2025CLIBD}. These methods are useful when the goal is not only to detect an organism, but also to compare it with a reference collection, retrieve similar examples, and use taxonomic metadata to support identification.

For marine species recognition, large curated datasets and benchmarks are essential because fine-grained visual differences between species are often subtle. FishNet provides a large-scale benchmark for fish recognition, detection, and trait prediction, supporting the development of models that are better aligned with marine biodiversity tasks than generic object-recognition datasets \cite{Khan2023FishNet}. A practical retrieval-based pipeline can therefore be built by embedding a labeled reference dataset of fish or marine organisms and storing the resulting vectors in a search index. Classifying a new observation is then performed via cosine similarity or $k$-nearest-neighbor search. The retrieved nearest neighbors yield both a predicted label and interpretable supporting evidence, such as visually similar reference images, taxonomic names, habitat metadata, or confidence estimates.

Multimodal retrieval-augmented generation extends this idea by combining retrieval with a generative or reasoning model. In such a system, image, text or acoustic embeddings retrieve relevant reference examples and metadata before a vision--language or multimodal language model produces a final explanation or report. Recent fisheries-oriented multimodal RAG work follows this direction by integrating heterogeneous fishery data sources for domain-specific retrieval and reasoning \citep{Song2026FMRAG}. For autonomous marine monitoring, this suggests a sequential approach : Filtering detections with low-consumption models, then generating species hypotheses and answers with LLM models.

\section{Datasets and Taxonomic Resources}
\label{sec:datasets}

This section groups all data sources used by the system. The first part describes the datasets used to train and validate the two low-power sentinel subsystems: visual fish detection and marine mammal acoustic classification. The second part describes the taxonomic reference collections used by the high-fidelity identification and retrieval layer after Jetson activation.

\subsection{FishDet-M Image Dataset}
\label{subsec:fishdetm}

The objective of the visual sentinel is to detect underwater events and then decide, according to the currently selected operating mode, whether the Jetson should be triggered or just store the data on the SD Card. As an application example, we restrict to fish detection. The visual models are trained on the FishDet-M dataset \citep{abujabal2025fishdetmunifiedlargescalebenchmark}.

FishDet-M is a large-scale, unified benchmark dataset designed for robust fish and underwater object detection. To address the historical challenge of fragmented and niche marine vision datasets, FishDet-M consolidates 13 distinct public underwater datasets, including well-known sources such as DeepFish, FishNet, Brackish-MOT, and TrashCan 1.0, into a single standardized framework. The resulting collection contains 105,556 images and 296,885 annotated fish instances, all unified under standard COCO-style annotations with both bounding boxes and segmentation masks. The dataset is explicitly curated to test domain generalization across highly diverse aquatic visual domains, ranging from tropical coral reefs and marine environments to brackish waters, aquaculture tanks, and indoor aquariums. As a result, models trained on this dataset are exposed to realistic deployment challenges such as high turbidity, severe occlusions, motion blur, muted contrast, and green-blue spectral distortion.

To enhance model generalization and prevent overfitting, a dynamic stochastic data augmentation strategy is used during training. For each sample, the pipeline dynamically generates two distinct versions by randomly applying transformations such as geometric modifications, color jittering, and noise insertion. We restricted the training and testing phases to images containing five or fewer bounding boxes. First, the models used are too compact to capture very fine details. Second, resizing the images to 256×320 pixels causes some bounding boxes to become excessively small and irrelevant. Finally, according to Tinyissimo Yolo model original paper, the authors gained approximately 20\% mAP by limiting detection to a maximum of five objects per frame \cite{Moosmann2023TinyissimoYOLO}. The final dataset repartition is resumed in table \ref{tab:dataset_stats}.

\begin{table}[t]
\centering
\caption{Overview of the FishDet-M dataset splits and annotations before and after restricting to five or fewer fish instances per image.}
\label{tab:dataset_stats}
\resizebox{\columnwidth}{!}{%
\begin{tabular}{lccc}
\toprule
\textbf{Feature} & \textbf{Training Set} & \textbf{Validation Set} & \textbf{Test Set} \\
\midrule
\multicolumn{4}{l}{\textit{Original Dataset (No Restrictions)}} \\
\midrule
Number of Images & 83{,}093 & 10{,}654 & 11{,}809 \\
Number of Fish Instances & 228{,}558 & 32{,}961 & 35{,}366 \\
\midrule
\multicolumn{4}{l}{\textit{Filtered Dataset ($\le$ 5 Fish Instances per Image)}} \\
\midrule
Number of Images & 76{,}202 & 9{,}626 & 10{,}645 \\
Number of Fish Instances & 95{,}199 & 12{,}290 & 14{,}060 \\
\bottomrule
\end{tabular}%
}
\end{table}

\subsection{Marine Mammal Acoustic Dataset}
\label{subsec:acoustic_dataset}

For the initial underwater acoustic experiments, the system focuses on marine mammal detection. Many marine mammals emit sounds in the frequency range from 200 Hz to 8,000 Hz, which aligns with the memory limitations of the MAX78000 microcontroller. To natively match the hardware capture behavior of the MAX9867 audio CODEC, all input audio is isolated to the left channel, resampled to a 16 kHz sample rate, and divided into 0.98-second frame windows.

The primary dataset selected for this task is the Watkins Marine Mammal Sound Database \citep{watkins_dataset}. This open-access historical archive contains thousands of digitized underwater audio recordings of more than 60 species of whales, dolphins, and seals, collected globally over seven decades. To prevent class imbalance from skewing model performance, the dataset is restricted to the 12 most heavily represented species.

To make the model robust against ambient underwater sounds, a \textbf{Noise} class is introduced as Class 0. The dataset used to represent this background noise is DeepShip \citep{IRFAN2021115270}, a benchmark underwater acoustics dataset consisting of more than 47 hours of real-world passive sonar recordings. DeepShip is commonly used to train machine learning models that classify four distinct commercial ship types: cargo, passenger, tanker, and tug.

\begin{table}[width=.9\linewidth,cols=3,pos=h]
\caption{Dataset distribution across classes. Number of elements and durations for Whale and Dolphin classes are sourced from the WHOI Dataset data exploration. Noise class is sourced from DeepShip.}\label{tab:dataset_summary}
\begin{tabular*}{\tblwidth}{@{} LLL@{} }
\toprule
Class Name & Audio files & Duration (h)\\
\midrule
Noise (DeepShip) & 63 & > 47.0 \\
Common Dolphin & 884 & 0.64 \\
False KillerWhale & 508 & 0.27 \\
FinbackWhale & 580 & 3.90 \\
HumpbackWhale & 604 & 1.99 \\
KillerWhale & 2647 & 1.65 \\
Long Finned PilotWhale & 1104 & 0.86 \\
Pantropical Spotted Dolphin & 1025 & 0.85 \\
Short Finned PilotWhale & 607 & 0.39 \\
SpermWhale & 1379 & 12.26 \\
Spinner Dolphin & 524 & 0.66 \\
Striped Dolphin & 681 & 0.36 \\
White sided Dolphin & 560 & 0.45 \\
\bottomrule
\end{tabular*}
\end{table}

The inherent static noise produced by the MAX9867 audio CODEC on the MAX78000 platform creates a third technical challenge. To address this issue and ensure the \textbf{Noise} class matches the volume of the most heavily represented species without overfitting, the dataset is dynamically balanced using real hardware baseline recordings combined with a three-pronged synthetic noise generation strategy:
\begin{enumerate}
    \item \textbf{Board Noise Crops:} Random 0.98-second windows are extracted from real MAX9867 background recordings and perturbed with Gaussian jitter ($\sigma=2$ on a 16-bit scale) to simulate varied board idle floors.
    \item \textbf{White Gaussian Noise:} Flat-spectrum noise generated at random amplitudes reflecting the typical board idle floor (-100 dBFS to -60 dBFS).
    \item \textbf{Pink (1/f) Noise:} Noise that rolls off at 3 dB/octave, generated between -90 dBFS and -55 dBFS, closely mimicking the ambient ocean spectral shape found in real underwater recordings.
\end{enumerate}

Prior to Mel spectrogram generation, all data (both real and synthetic) is scaled to a 16-bit PCM integer range and passed through an exact 1st-order IIR high-pass filter (coefficients $b = [1.0, -1.0]$, $a = [1.0, -0.995]$) to perfectly mirror the C-level firmware execution on the hardware.

\begin{figure}
    \centering
    \includegraphics[width=1\linewidth]{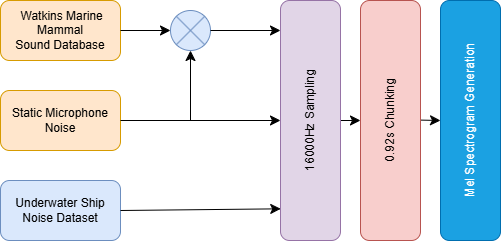}
    \caption{Acoustic data generation pipeline.}
    \label{fig:acoustic_dataset}
\end{figure}

The aggregated dataset is ultimately partitioned into a 70\% training, 20\% validation, and 10\% testing split. By handling hardware noise and signal filtering during the training phase, the system avoids the need for complex and computationally expensive noise-filtering steps during real-time preprocessing. This minimizes both latency and energy consumption. After merging all datasets, a standardized pipeline is applied to generate 96x64 log-Mel spectrograms. This pipeline is strictly designed to remain memory efficient, because the same preprocessing steps must be executed locally on the MAX78000 in real time, while avoiding biases that could lead to misclassification. Figure~\ref{fig:acoustic_dataset} summarizes the complete acoustic dataset generation process.

\subsection{Taxonomic Reference Collections and Ingestion}
\label{subsec:taxonomic_datasets_prep}

The high-fidelity identification layer relies on extensive taxonomic reference collections to cross-reference sensor observations. As summarized in Table~\ref{tab:datasets}, the project leverages five indexed reference datasets—FishBase/SeaLifeBase \cite{fishbase}\cite{sealifebase}, BioTrove\cite{biotrove}, FishNet \cite{fishnet}, Fish-Vista\cite{fishvista}, and WildFish++ \cite{WildFishPlusPlus2018} —encompassing 493,387 visual entries in total. OzFish \cite{AIMS_OzFish2020} is intentionally excluded from the reference index and reserved strictly as an external evaluation set. 

Table \ref{tab:datasets} summarizes the different taxonomic reference datasets by specifying the number of images, species, genera, families, orders, and classes, as well as their geographic or ecological scope. 

\begin{table*}[ht]
\small
\caption{Taxonomic image collections used by the project. Image counts correspond to indexed visual entries when the collection is used as a reference. OzFish is reported separately because it 
is used only for evaluation. Taxonomic counts are computed after the project normalization step and, when available, completed with the local taxonomy cache.}\label{tab:datasets}
\begin{tabular*}{\tblwidth}{@{} p{1.5cm} p{1cm} p{1cm} p{1cm} p{1cm} p{1cm} p{1cm} p{7.0cm} @{} }
\toprule
Dataset & Images & Species & Genera & Families & Orders & Classes & Image conditions\\
\midrule
FishBase / \\SeaLifeBase & 49,149 & 17,379 & 5,489 & 1,436 & 213 & 60 & Reference photographs from curated databases; mixed live, aquarium, field, specimen, and preserved/dead photographs 
depending on source metadata. \\
BioTrove & 190,534 & 19,586 & 5,739 & 1,074 & 188 & 35 & Heterogeneous reference images organized by taxonomic class and species; conditions are mixed rather than exclusively in situ. \\
FishNet & 94,349 & 17,610 & 3,973 & 581 & 81 & 10 & Mostly object/reference-style fish images, including multiple life stages and non-field compositions; not a pure underwater in situ dataset. 
\\
Fish-Vista & 56,321 & 729 & 309 & 120 & 47 & 3 & Curated fish photographs from GLIN, iDigBio, and MorphBank; mixed specimen, museum, and field-like images rather than only live underwater 
scenes. \\
WildFish++ & 103,034 & 2,348 & 931 & 255 & 62 & 7 & Real-world wild fish imagery from train and validation folders, closer to natural visual conditions than the database reference collections. 
\\
OzFish & 1,903 & 198 & 84 & 29 & 10 & 2 & Real fish observations with species boxes and measurement files; used to test cross-source generalization. \\
\bottomrule
\end{tabular*}
\end{table*}
\section{System Architecture}
\label{sec:system_architecture}

This section describes the proposed underwater monitoring system architecture. The system is designed around a hierarchical edge-computing structure in which low-power sentinel microcontrollers continuously monitor the environment, while an NVIDIA Jetson Orin NX is activated only when higher-capacity multimodal processing is required. The section first introduces the hardware hierarchy and the power-aware operating modes. It then describes the local multimodal processing pipeline executed on the Jetson Orin NX, including detection-aware ingestion, visual target extraction, multimodal embedding, species identification, and retrieval-augmented reasoning. Finally, it presents the multi-agent control framework responsible for routing, analysis, reporting, energy management, and hardware adaptation.

\subsection{Hierarchical Structural Overview}
\label{subsec:structural_overview}

The proposed underwater monitoring system follows a hierarchical master--satellite architecture designed to reconcile continuous environmental monitoring with strict energy constraints. The architecture separates always-on low-power perception from high-performance multimodal reasoning.

At the lower tier, a MAX78000 microcontroller and another MAX78002 microcontroller operate as low-power sentinel nodes dedicated to acoustic and visual monitoring. These subsystems continuously execute lightweight neural-network models to detect potentially relevant events, such as marine mammal vocalizations or visual activity. Their role is not to perform complete scientific interpretation, but to provide an energy-efficient first filtering stage that determines whether an event deserves deeper analysis.

At the upper tier, an NVIDIA Jetson Orin NX acts as the main computational hub. It is responsible for high-fidelity inference, multimodal embedding, local vector indexing, species identification, retrieval-augmented generation, researcher interaction, reporting, and agentic orchestration. Under normal conditions, the Jetson remains shut down and is activated only when scheduled processing, event-driven analysis, or researcher interaction is required.

This division of labor allows the system to preserve energy during long-duration underwater deployments while still providing access to computationally intensive AI pipelines when the collected data justify deeper analysis.

\begin{figure*}
\centering
\includegraphics[width=1\linewidth]{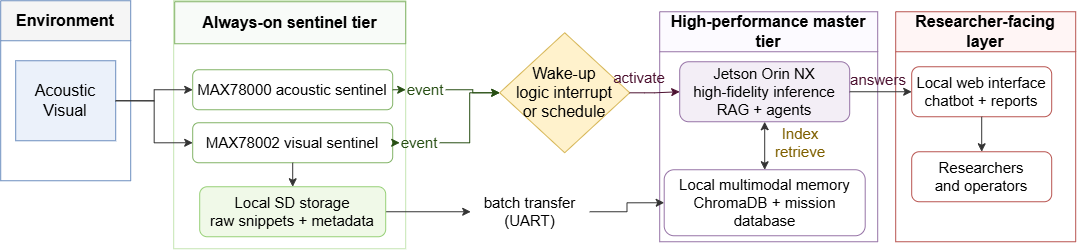}
\caption{Hierarchical, low-power underwater monitoring architecture. Edge-level microcontrollers ($\text{MAX78000/02}$) continuously monitor environmental events and selectively wake a high-performance master node ($\text{Jetson Orin NX}$) for local multimodal processing, retrieval-augmented generation, and researcher interaction.}
\label{fig:structural_overview}
\end{figure*}

\begin{figure}
\centering
\includegraphics[width=\linewidth]{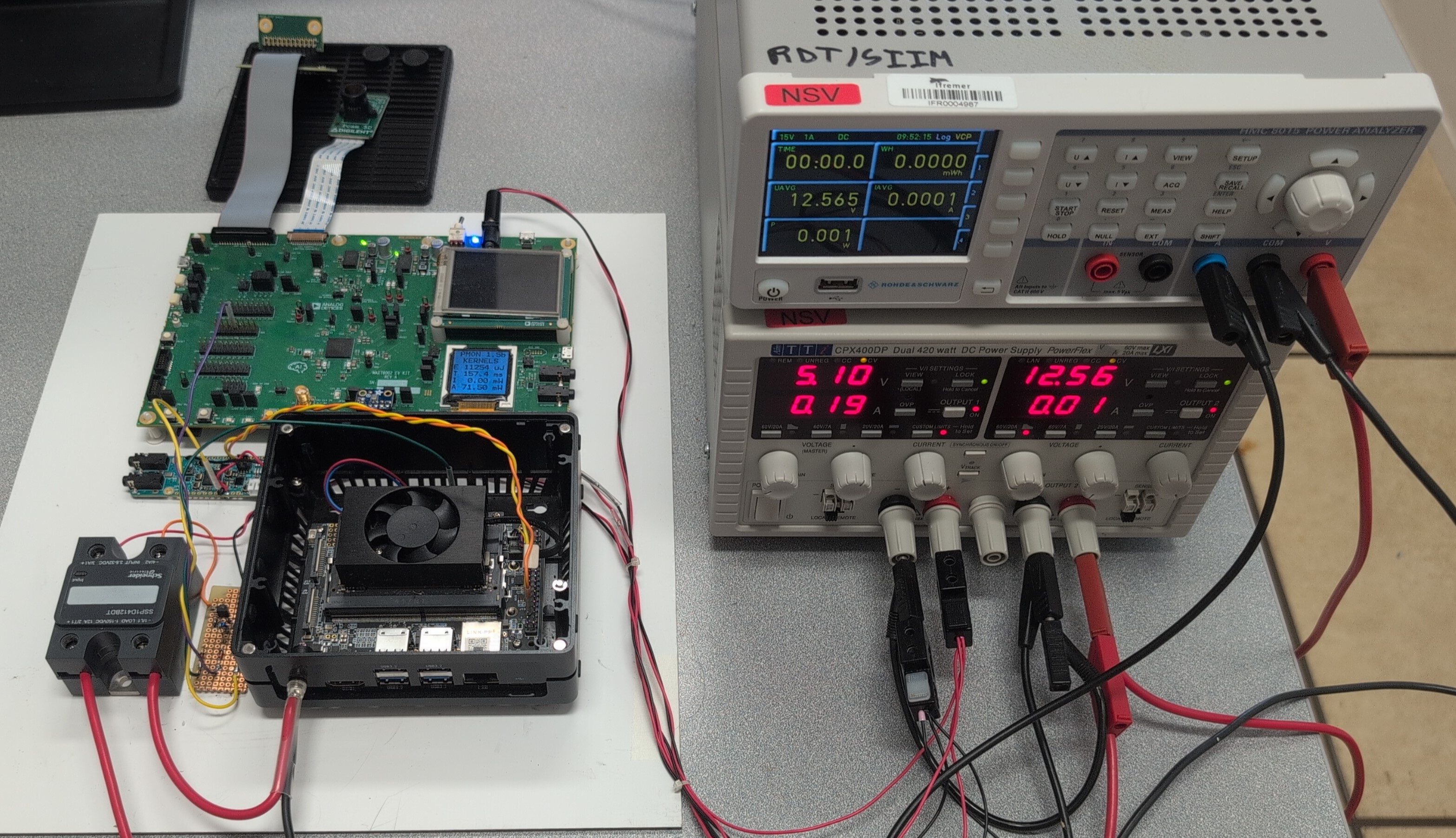}
\caption{Experimental implementation of the system}
\label{fig:experimental_system_A}
\end{figure}

Figure \ref{fig:structural_overview} resumes all the points mentionned in this section. Figure \ref{fig:experimental_system_A} shows a prototype of the developed system. 
\subsection{Power-Aware Operation and Hardware Orchestration}
\label{subsec:power_aware_operation}

The system supports three operating modes, each corresponding to a different trade-off between autonomy, latency, power consumption, and researcher interaction. In all modes, the low-power sentinel tier remains responsible for continuous environmental monitoring, while the Jetson Orin NX is activated selectively.

\subsubsection{Autonomous Batch Mode}
\label{subsubsec:autonomous_batch_mode}

Autonomous Batch Mode is intended for long-duration deployments where energy preservation is the main constraint. The MAX-based acoustic and visual subsystems remain active and locally store raw detections on their respective SD cards. 

In this mode, the Jetson Orin NX remains turned off most of the time and wakes according to a scheduled cycle, for example once per day. During this scheduled activation, the Jetson ingests the accumulated detections, transfers the corresponding files from the sentinel storage arrays through UART, performs secondary validation using higher-capacity models, indexes the new data, generates compact scientific summaries, and then returns to a low-power state.

\subsubsection{Event-Driven Hybrid Mode}
\label{subsubsec:event_driven_hybrid_mode}

Event-Driven Hybrid Mode prioritizes low-latency situational awareness while preserving the sentinel-first architecture. When a low-power subsystem detects an event of interest, it immediately wakes the Jetson Orin NX through the hardware orchestration layer. The Jetson can then retrieve the corresponding raw data, activate additional sensors if necessary, perform deeper multimodal analysis, and update the local database shortly after the event occurs.

This mode is suitable for deployments in which relevant biological or environmental events must be analyzed soon after detection, while continuous operation of the high-power compute module remains undesirable.

\begin{figure}
    \centering
    \includegraphics[width=\linewidth]{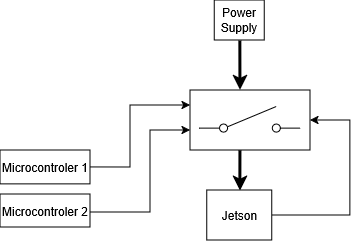}
    \caption{Hardware wake-up and power-hold circuit used in Event-Driven Hybrid Mode.}
    \label{fig:electric_circuit}
\end{figure}

To implement this behavior, the event detection sets microcontrollers' I/O that are connected to a shared \textit{TTL} wake line, as illustrated in Fig.~\ref{fig:electric_circuit}. This simple diode-OR configuration allows any microcontroller to initiate a wake-up event while preventing current from flowing between GPIO outputs. Assertion of the shared event line activates the hardware power-switching stage, which connects the \(12\,\mathrm{V}\) supply to the Jetson.

After booting, the Jetson asserts a dedicated GPIO connected to the same control node through an additional isolation diode. This signal provides a self-hold function, maintaining power after the initiating event signal has been released. Once event processing, data transfer, and database updates are complete, the Jetson performs an orderly software shutdown and releases the hold signal, causing the power-switching stage to disconnect the \(12\,\mathrm{V}\) supply. Consequently, short detection pulses are sufficient to initiate a complete processing cycle without requiring the detecting microcontroller to remain active throughout the Jetson operating period.

\subsubsection{Mission and Researcher-Interactive Mode}
\label{subsubsec:mission_mode}

Mission and Researcher-Interactive Mode is designed for field operations in which researchers are physically present near the monitoring station and require direct access to the local intelligence layer. In this mode, the Jetson Orin NX remains active and exposes a local web interface implemented using HTML, CSS, and JavaScript. Unlike Autonomous Batch Mode and Event-Driven Hybrid Mode, this mode prioritizes interactive exploration, rapid scientific verification, and direct access to indexed data over minimum power consumption.

The interface is organized into four main tabs. The first tab provides a chatbot interface connected to the local agentic retrieval system. Researchers can submit natural-language queries, which are processed by the backend through the local multimodal RAG pipeline and the unified agentic reasoning layer. Depending on the query, the system routes the request toward the appropriate ChromaDB collection, retrieval modality, analytical pathway, or local instruct model.

\begin{figure}
    \centering
    \includegraphics[width=1\linewidth]{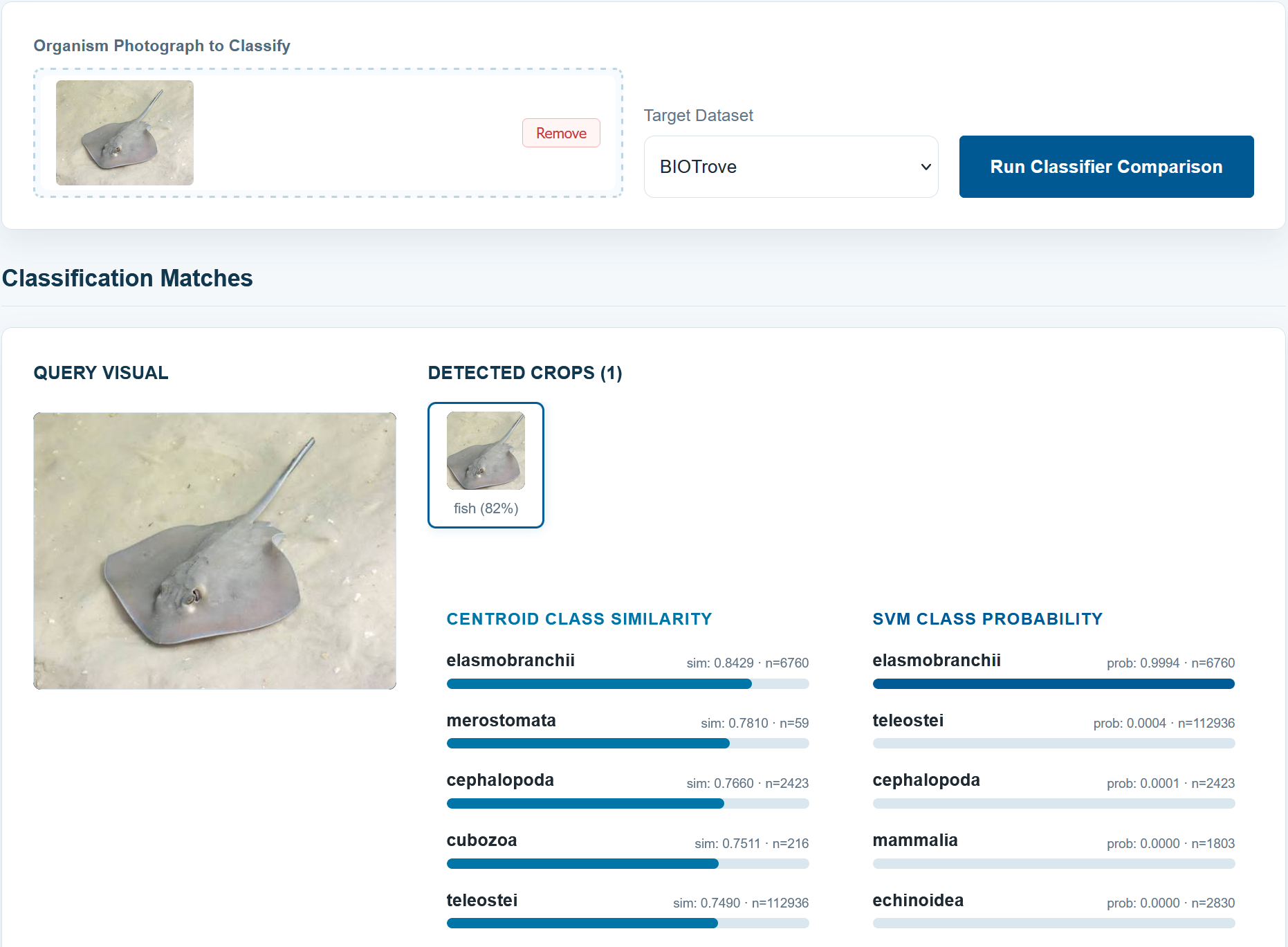}
    \caption{Specie identification web interface}
    \label{fig:identification_layer}
\end{figure}

\begin{figure*}
    \centering
    \includegraphics[width=1\linewidth]{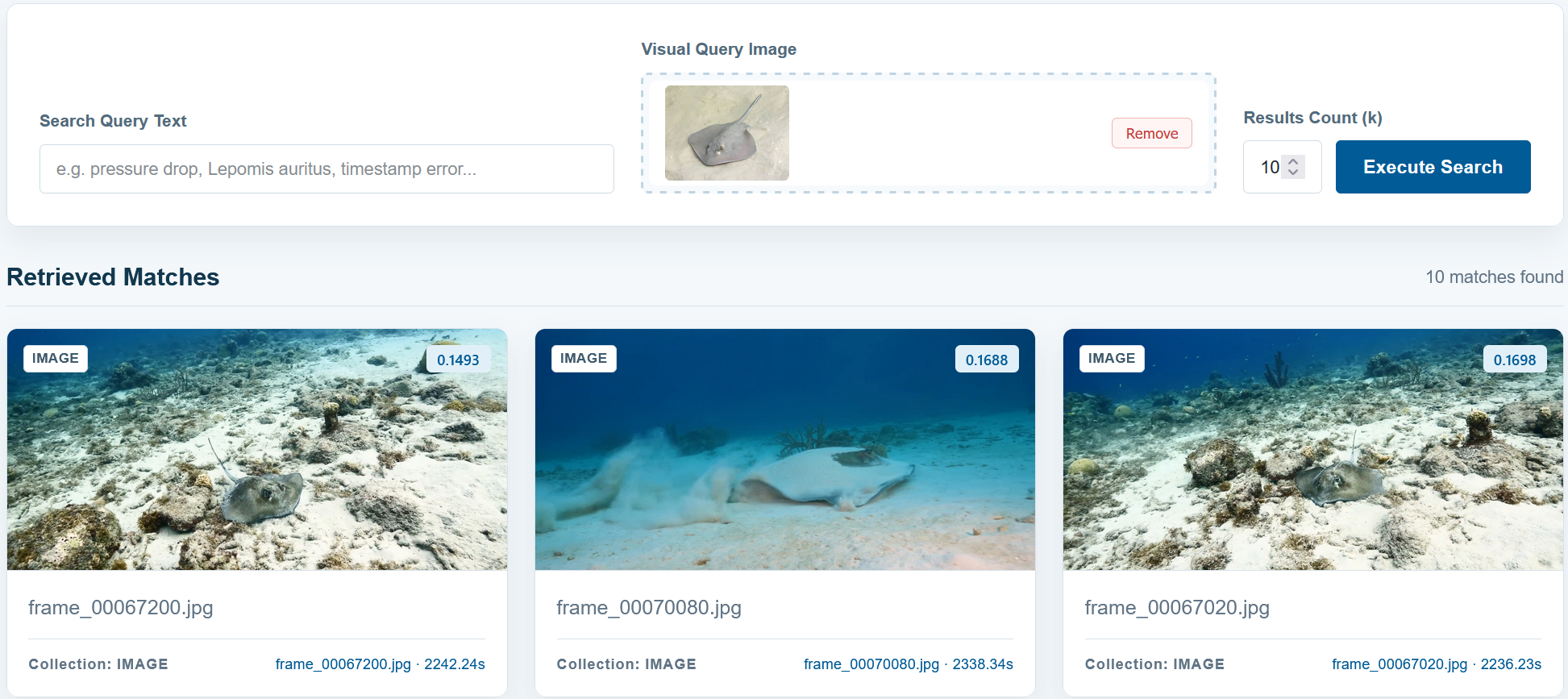}
    \caption{Example of image-image similarity search using the Web UI}
    \label{fig:image_similarity_search}
\end{figure*}

\begin{figure}
    \centering
    \includegraphics[width=1\linewidth]{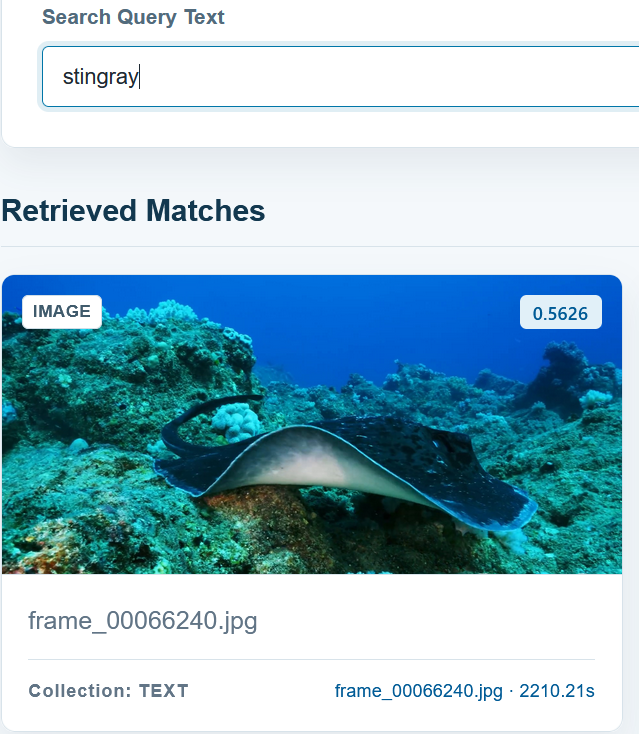}
    \caption{Example of text-image similarity search}
    \label{fig:text_similarity_search}
\end{figure}

The second tab is dedicated to species identification. A researcher can upload an image of an unknown organism and select the reference collection to be used for identification. The backend then compares the image embedding against precomputed taxonomic centroids and SVM classifiers. The interface returns the most likely species, genus, family, order, and class, together with the corresponding confidence indicators. For SVM-based predictions, the reported value corresponds to the classifier probability, while for centroid-based predictions it corresponds to the cosine similarity between the query embedding and the selected taxonomic centroid. Figure \ref{fig:identification_layer} shows a specie identification example, the Web UI displays the five taxonomic levels (class, order, genus, family and specie).
The third tab supports visual similarity search. The researcher can upload or select a query image, describe a text query, choose a target collection, and define the number of nearest neighbors $k$ to retrieve. The search can be performed either over marine taxonomic reference datasets or over the local multimodal mission collection. The interface then displays the top-$k$ closest images together with their similarity scores, allowing researchers to visually inspect related examples, compare uncertain detections, and validate retrieval results. Figure \ref{fig:image_similarity_search} shows a similarity search on image, and figure \ref{fig:text_similarity_search} shows a similarity search but this time on text.

The fourth tab provides a database overview. It lists the currently available ChromaDB collections and reports the number of indexed items in each collection. This view allows researchers to verify the state of the local memory, inspect whether new detections or reference assets have been indexed, and monitor the growth of the onboard database during field operations.

Overall, Mission and Researcher-Interactive Mode turns the monitoring station into a local scientific inspection platform. It allows researchers to query the RAG system, identify species from uploaded images, retrieve visually similar examples, and inspect the current database state directly at the edge without requiring cloud connectivity.

\subsection{Local Multimodal Processing Pipeline}
\label{subsec:local_multimodal_pipeline}

Once the Jetson Orin NX is activated, the collected data are processed through a fully local multimodal pipeline. This pipeline transforms raw detections into indexed, searchable, and interpretable scientific information. It links real-time detections with historical mission data, taxonomic references, and scientific documentation without requiring cloud connectivity.

The pipeline consists of four main stages: (i) detection-aware data ingestion, (ii) target localization, (iii) species identification, multimodal embedding, and indexing, and (iv) retrieval-augmented reasoning. This modular organization clearly separates data management from species identification, making the role of each stage explicit and facilitating future extensions.

Before the pipeline can be executed, the species identification layer must be initialized. During this configuration stage, a set of reference taxonomic datasets is indexed, after which class centroids and Support Vector Machine (SVM) classifiers are computed. These classifiers are subsequently used by the online multimodal pipeline to generate species hypotheses. The species identification layer is described in detail in Section \ref{subsubsec:species_identification_layer}.

The overall workflow is as follows. An event detected by the Sentinel system triggers the acquisition process. Depending on the operating mode, the acquired data (images or audio recordings) are either stored on the SD card for subsequent processing or captured directly after the Jetson Orin NX is awakened from its low-power state. The target localization stage extracts the relevant regions of interest from the acquired images using text-guided localization prompts. These image crops are then forwarded to the embedding and indexing stage, where BIOCLIP-2 embeddings are computed. The resulting embeddings are subsequently processed by the species identification layer to generate one or more hypotheses regarding the detected species.

The computed embeddings, together with their associated metadata—including timestamps, species hypotheses, mission identifiers, sensor information, and video sources are stored in ChromaDB collections. These indexed representations form a searchable knowledge base that continuously grows throughout the mission. When a query is submitted by either a human operator or an autonomous agent, the system performs retrieval-augmented generation (RAG) over the indexed ChromaDB collections. The retrieved embeddings and their associated metadata are provided as contextual information to the large language model (LLM), enabling scientifically grounded responses based on both the current observations and previously collected mission data.

The following sections describe each stage of the proposed pipeline in detail. The presentation of the species identification layer is intentionally deferred until after the embedding stage, as it relies on embedding representations introduced in the preceding section.

\begin{figure*}
\centering
\includegraphics[width=1\linewidth]{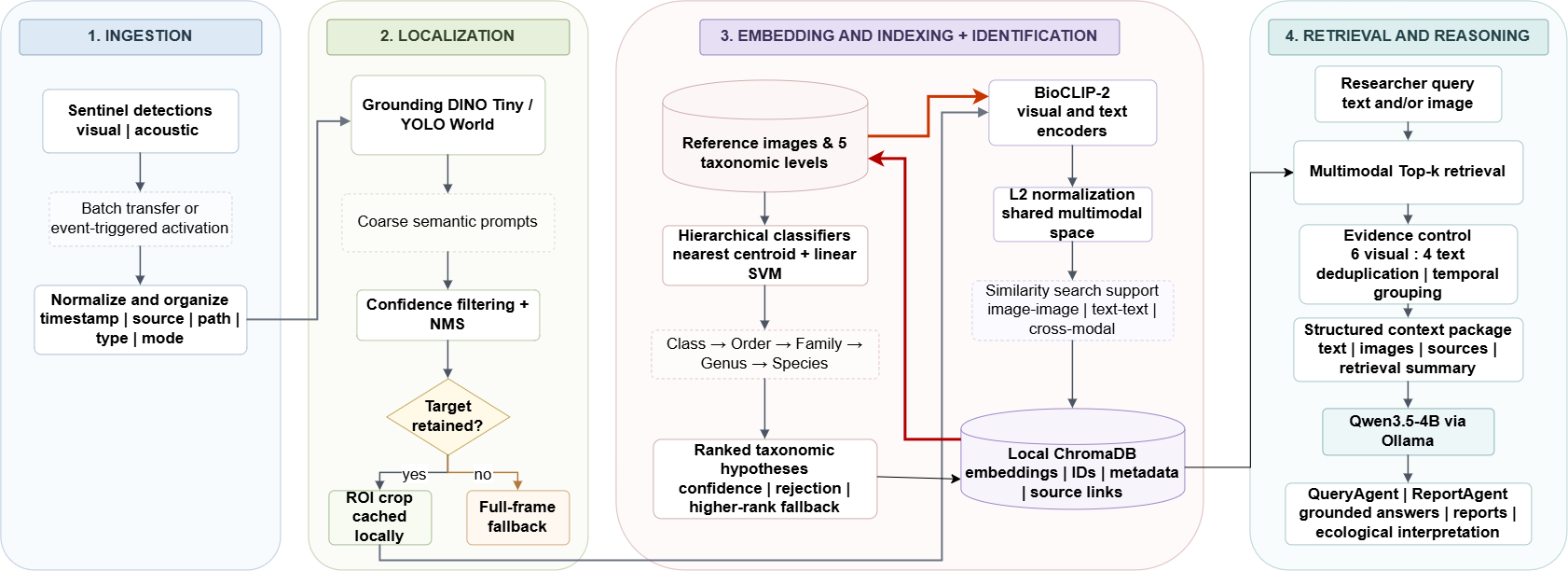}
\caption{Local multimodal processing and retrieval-augmented memory pipeline.}
\label{fig:multimodal_rag_pipeline}
\end{figure*}

\subsubsection{Detection-Aware Data Ingestion}
\label{subsubsec:detection_aware_ingestion}

The pipeline begins with detections produced by the low-power sentinel subsystems. These detections may include acoustic frames, seismic waves, visual observations or associated metadata. In Autonomous Batch Mode, detections are accumulated locally and transferred through UART to the Jetson during scheduled wake-up cycles. In Event-Driven Hybrid Mode, relevant detections can trigger immediate Jetson activation and data transfer.

At this stage, the objective is not yet to identify species or generate ecological explanations. The goal is to collect, normalize, timestamp, and organize the data produced by the sentinel layer so that they can be processed by the higher-capacity multimodal pipeline. Each ingested item is associated with metadata such as source subsystem, timestamp, file path, detection type, and triggering mode.

\subsubsection{Target Localization and Crop Extraction}
\label{subsubsec:target_localization}

For visual inputs, the pipeline adopts a detection-first strategy prior to vector database insertion, as illustrated in 
Figure~\ref{fig:multimodal_rag_pipeline}. The objective of this stage is to isolate semantically relevant marine organisms from the surrounding 
scene so that the subsequent embedding process focuses primarily on the target rather than on potentially dominant background elements, such as 
water, rocks, sand, vegetation, or acquisition equipment.

Target localization is performed using Grounding DINO \cite{liu2024groundingdinomarryingdino}, an open-vocabulary object detector that combines 
the transformer-based DINO detection architecture with grounded vision--language pre-training. Unlike conventional closed-set detectors, which 
are restricted to a predefined collection of object classes, Grounding DINO receives natural-language expressions as detection queries. It 
jointly encodes the input image and the textual prompt, aligns their visual and linguistic representations, and returns bounding boxes 
associated with the image regions that are most semantically consistent with the requested concepts. This property is particularly suitable for 
marine imagery, for which the taxonomic categories of interest may vary between deployments and may not be adequately represented by standard 
object-detection datasets.

The pipeline uses the Grounding DINO Tiny variant, based on a Swin-T visual backbone. Despite the ``Tiny'' designation, the complete model 
contains approximately \(172\) million parameters. The storage required by the parameters alone is therefore approximately \(688\)~MB when 
represented in 32-bit floating-point precision and approximately \(344\)~MB under 16-bit precision. These values represent only the theoretical 
memory required to store the model weights. During inference, additional GPU memory is required for visual feature maps, language 
representations, cross-modal attention tensors, decoder queries, intermediate activations, and framework-level CUDA allocations. Consequently, 
several gigabytes of GPU memory must be reserved in practice; an allocation of approximately \(4\)~GB provides a reasonable operating margin 
for single-image inference, although the precise peak consumption depends on the input resolution, numerical precision, software 
implementation, and memory-caching behavior.

At inference time, the detector is queried using coarse semantic categories such as \textit{fish}, \textit{cnidarians}, \textit{mollusks}, and 
\textit{crustaceans}. These prompts are intentionally broader than species-level labels because the purpose of this stage is target 
localization rather than fine-grained taxonomic identification. For every input frame, Grounding DINO produces a collection of candidate 
bounding boxes and their corresponding confidence scores. Low-confidence predictions are rejected, after which Non-Maximum Suppression (NMS) is 
applied to remove strongly overlapping detections that are likely to correspond to the same organism. The remaining regions of interest are 
cropped from the original frame and cached locally, thereby avoiding repeated detection and crop-extraction operations during subsequent 
indexing or retrieval experiments.

As summarized in Figure~\ref{fig:multimodal_rag_pipeline}, the retained crops are subsequently passed to the visual embedding model and 
inserted into the vector database together with their associated metadata. Performing detection before feature extraction reduces the influence 
of irrelevant background content and produces embeddings that more directly represent the visual characteristics of the detected organism. This 
is especially important when the target occupies only a small portion of the original image or when multiple background structures exhibit 
stronger visual patterns than the organism itself.

If no candidate satisfies the detection-confidence criteria, the complete frame is retained as a fallback representation. This conservative 
mechanism prevents potentially informative observations from being discarded due to difficult imaging conditions, small or partially occluded 
organisms, domain shifts, or imperfect agreement between the selected textual prompts and the visible target. Such samples can therefore remain 
available for downstream identification, retrieval, and anomaly analysis even when reliable localization is not achieved.

Among the models involved in the indexing pipeline, Grounding DINO is the most computationally expensive component. Its computational cost 
arises from the simultaneous processing of high-resolution visual features and textual representations, together with repeated cross-modal 
attention and bounding-box decoding operations. In comparison, crop extraction, NMS, local caching, and vector-database insertion introduce 
relatively limited overhead. Grounding DINO consequently accounts for the largest portion of the per-frame processing time and constitutes the 
principal computational bottleneck of the detection-aware indexing procedure. This motivates executing localization only once during ingestion 
and caching its outputs, rather than repeating detection whenever an indexed observation is queried. The final output of this stage is a set of 
localized candidate regions---or the original frame when localization fails---that can subsequently be embedded, indexed, identified, and 
retrieved.

We also evaluated YOLO World \cite{lu2024yoloworld}, which achieved very similar detection performance to Grounding DINO while operating at substantially lower computational cost. Specifically, YOLO World completes inference in a fraction of the time required by Grounding DINO for equivalent tasks, making it well-suited for real-time or near-real-time applications where throughput is critical. Despite this speed advantage, its semantic grounding capability and open-vocabulary detection accuracy remain comparable to those of Grounding DINO, ensuring that neither approach introduces meaningful differences in downstream retrieval quality. Yolo World may also be deployed through TensorRT which makes the inference far faster specially on Jetson devices.
\subsubsection{Multimodal Embedding and Vector Indexing}
\label{subsubsec:embedding_indexing}

After candidate data have been collected and the relevant visual regions have been extracted, the pipeline converts each element into a fixed-dimensional numerical representation, referred to as an \emph{embedding}. These embeddings encode the semantic content of an input and provide a common representation in which heterogeneous modalities can be compared and retrieved efficiently.

The embedding stage utilizes Contrastive Language--Image Pre-training (CLIP) \cite{Radford2021CLIP}, a multimodal architecture featuring separate visual and text encoders. By mapping matching image--text pairs into a shared embedding space during training, CLIP aligns semantically related visual and textual representations, enabling direct cross-modal comparison without a task-specific classification head.

In the proposed pipeline, visual inputs---including detected image crops and sampled video frames---are processed using either BioCLIP-2 or a general-purpose OpenCLIP visual encoder. OpenCLIP is an open-source implementation of the CLIP architecture that provides pretrained models with different visual backbones, training datasets, and computational requirements. It offers a general representation of visual and linguistic concepts learned from large collections of image--text pairs. However, because these training collections primarily contain broad web imagery, the resulting representation may provide limited discrimination between visually similar biological taxa.

BioCLIP \cite{Stevens2024BioCLIP} addresses this limitation by adapting the CLIP training paradigm to the biological domain. It retains the dual-encoder architecture and contrastive learning objective of CLIP but is trained using TreeOfLife-10M, a large-scale dataset containing images of animals, plants, and fungi together with structured taxonomic labels. The textual supervision incorporates the biological hierarchy, including taxonomic ranks such as kingdom, phylum, class, order, family, genus, and species. As a result, the learned embedding space captures both visual similarity and biologically meaningful relationships between organisms. This domain specialization makes BioCLIP more appropriate than a general-purpose CLIP model for representing marine species, particularly when the downstream task requires fine-grained differentiation between morphologically similar taxa.

For each visual observation \(I_i\), the visual encoder produces an embedding vector
\begin{equation}
    \mathbf{v}_i = f_{\mathrm{img}}(I_i),
\end{equation}
where \(f_{\mathrm{img}}\) denotes the BioCLIP or OpenCLIP image encoder. Similarly, each textual element \(T_j\), such as a scientific name, species description, ecological trait, document chunk, or operational annotation, is transformed using the corresponding text encoder:
\begin{equation}
    \mathbf{t}_j = f_{\mathrm{text}}(T_j).
\end{equation}

Before indexing, the resulting vectors are \(L_2\)-normalized:
\begin{equation}
    \hat{\mathbf{z}} =
    \frac{\mathbf{z}}{\lVert \mathbf{z} \rVert_2},
\end{equation}
where \(\mathbf{z}\) represents either a visual or textual embedding. Semantic similarity between two normalized representations is subsequently measured using cosine similarity, which reduces to their dot product:
\begin{equation}
    s(\hat{\mathbf{z}}_a,\hat{\mathbf{z}}_b)
    =
    \hat{\mathbf{z}}_a^{\top}\hat{\mathbf{z}}_b.
\end{equation}
A larger similarity value indicates that the corresponding inputs are more closely related within the learned multimodal space. This formulation supports image-to-image, text-to-text, and cross-modal text-to-image retrieval using the same indexing mechanism.

The computational cost and memory footprint of this stage depend on the selected CLIP backbone. Models based on larger Vision Transformer architectures generally provide richer representations but require more parameters, GPU memory, and inference time. In the proposed pipeline, embedding extraction is significantly less computationally expensive than the preceding Grounding DINO localization stage \cite{liu2024groundingdinomarryingdino} and exhibits a computational footprint highly comparable to YOLO-World \cite{lu2024yoloworld}. Nevertheless, processing a large number of crops or video frames can produce a substantial cumulative cost, as image scenes containing a high density of detections yield a proportionally higher number of localized crops requiring embedding generation. Embeddings are therefore computed once during ingestion, stored persistently, and reused during subsequent retrieval and identification operations.

Mission images, localized video crops, textual species descriptions, PDF chunks and operational metadata are stored in modality-specific local ChromaDB collections. ChromaDB provides a vector-oriented storage layer that associates each embedding with a unique identifier, its original content or file reference, and structured metadata. Depending on the data modality, the stored metadata include the source-file path, frame or document index, acquisition timestamp, detected category, bounding-box coordinates, mission identifier, and preprocessing configurations. Crucially, the metadata also encompass hypotheses regarding taxonomic classification, which are detailed in the subsequent subsection. Preserving this provenance ensures that each retrieved vector remains traceable to its corresponding observation, thereby preventing the embedding space from decoupling from the original mission data. When multiple crops within a single image are assumed to belong to the same species, only one representative crop is retained. The total number of such instances is recorded within the metadata of the retained crop. This mechanism prevents the saturation of ChromaDB with redundant embeddings of the same species (e.g., a large school of sardines) while simultaneously preserving the original instance count.

During retrieval, a query is encoded using the appropriate pathway and compared with the indexed vectors. ChromaDB returns the nearest candidates according to their vector distance or similarity. For example, a textual query describing a marine organism can retrieve visually compatible image crops, while an image crop can retrieve related species descriptions or visually similar observations.

As previously noted, indexing the ingested data entails storing hypotheses across five distinct taxonomic levels as metadata. The methodology for generating these classifications is entirely delegated to the species identification layer, which is detailed in the subsequent subsection.

\subsubsection{Species Identification Layer}
\label{subsubsec:species_identification_layer}

Although multimodal embeddings provide a unified representation for images and text, similarity within this space does not inherently constitute a definitive taxonomic decision. Consequently, a dedicated species identification layer is introduced prior to the data indexing layer. This layer utilizes embeddings from reference sets to generate a ranked list of taxonomic hypotheses for each detected crop, accompanied by confidence scores and supporting evidence.

The architecture operates directly in the pretrained BioCLIP -2 embedding space, avoiding retraining for every mission or geographic scope. Mission adaptation is achieved by updating reference collections rather than maintaining multiple separately trained CLIP models: a coastal survey can use fish-specific references, while a broader biodiversity survey incorporates mollusks, cnidarians, and crustaceans alongside fish.

Each reference image is embedded together with a taxonomically structured textual representation. For an organism with known taxonomy, the text prompt follows the template
\begin{equation}
    T =
    \text{``a photo of }
    \langle\text{class}\rangle\;
    \langle\text{order}\rangle\;
    \langle\text{family}\rangle\;
    \langle\text{genus}\rangle\;
    \langle\text{species}\rangle
    \text{''},
\end{equation}
where every placeholder is replaced by its authoritative taxonomic value. The structured prompt provides the text encoder with both fine-grained species identity and hierarchical context.

Because source reference collections do not always provide all five taxonomic ranks, missing genus, order, or class values are reconstructed through a taxonomy-resolution infrastructure. Name normalization and taxonomic completion are supported by a synonym cache and a project-level taxonomy cache.

Let \(I_r\) denote a reference image and \(T_r\) its completed taxonomic prompt. Their normalized BioCLIP representations are
\begin{equation}
    \hat{\mathbf{v}}_r =
    \frac{f_{\mathrm{img}}(I_r)}
    {\lVert f_{\mathrm{img}}(I_r) \rVert_2},
    \qquad
    \hat{\mathbf{t}}_r =
    \frac{f_{\mathrm{text}}(T_r)}
    {\lVert f_{\mathrm{text}}(T_r) \rVert_2}.
\end{equation}
Visual and textual embeddings are stored as separate records in the same vector collection, preserving modality-specific information for independent evaluation during retrieval.

Given a candidate crop \(I_q\), its normalized BioCLIP visual embedding is
\begin{equation}
    \hat{\mathbf{v}}_q =
    \frac{f_{\mathrm{img}}(I_q)}
    {\left\lVert f_{\mathrm{img}}(I_q) \right\rVert_2}.
\end{equation}
Taxonomic prediction is performed using two complementary classifiers operating on the frozen BioCLIP embeddings.

For a taxon \(c\) represented by \(N_c\) reference image embeddings, the centroid is computed as
\begin{equation}
    \boldsymbol{\mu}_c =
    \frac{
        \sum_{i=1}^{N_c} \hat{\mathbf{v}}_{c,i}
    }{
        \left\lVert
        \sum_{i=1}^{N_c} \hat{\mathbf{v}}_{c,i}
        \right\rVert_2
    },
\end{equation}
where \(\hat{\mathbf{v}}_{c,i}\) is the normalized embedding of the \(i\)-th reference image for taxon \(c\). The query crop is compared with each cached centroid using cosine similarity:
\begin{equation}
    s_{\mathrm{centroid}}(q,c)
    =
    \hat{\mathbf{v}}_q^{\top}\boldsymbol{\mu}_c.
\end{equation}
Averaging over multiple reference observations reduces sensitivity to atypical viewpoints, noise, and annotation errors. Centroids are constructed at class, order, family, genus, and species levels, enabling hierarchical fallback when evidence is insufficient for a reliable species-level decision.

The second strategy uses multiclass linear Support Vector Machines trained on cached visual embeddings. For each taxonomic level \(l\), the score vector
\begin{equation}
    \mathbf{z}^{(l)}
    =
    \mathbf{W}^{(l)}\hat{\mathbf{v}}_q
    +
    \mathbf{b}^{(l)},
\end{equation}
is converted to normalized confidence via softmax:
\begin{equation}
    p^{(l)}_j
    =
    \frac{\exp\left(z^{(l)}_j\right)}{\sum_k \exp\left(z^{(l)}_k\right)}.
\end{equation}

Centroid and SVM classifiers provide complementary evidence: centroids preserve the geometric structure of the BioCLIP embedding space and are straightforward to update, while linear SVMs learn explicit decision boundaries that improve discrimination between visually similar taxa. Both operate on frozen embeddings without retraining the multimodal model.

The final identification decision combines both classifiers. Predictions below a configurable threshold are rejected; when species-level evidence is insufficient but higher-rank candidates agree, a broader taxonomic label (genus, family) is retained instead of forcing an unreliable species assignment. When multiple crops from the same frame receive identical predicted labels, only the top-\(N\) by score are preserved to reduce visual redundancy before downstream indexing. Each validated crop is stored with metadata (specie, genus, family, order, class, score, source file and acquisition timestamp).

The identification layer thus converts visual embeddings into traceable taxonomic hypotheses. Classification results are subsequently used by the retrieval-augmented reasoning layer to gather supporting biological, visual, temporal, and operational evidence from the local knowledge base.

Finally, the species identification layer can also be accessed via the Web user interface (UI). In Mission Mode, users can directly upload an image of marine fauna for classification. The interface then returns predictions from both the Support Vector Machine (SVM) and centroid-based models across five taxonomic levels. Figure \ref{fig:identification_layer} illustrates the species identification interface. For brevity, the figure displays predictions at the class level only; however, the full output provides predictions for all five taxonomic levels. This functionality provides researchers with immediate taxonomic hypotheses while maintaining a simple, user-friendly workflow.

\subsubsection{Retrieval-Augmented Reasoning (RAG)}
\label{subsubsec:retrieval_augmented_reasoning}

Once taxonomic candidates, event labels, or mission observations have been produced, the retrieval layer collects the evidence required by agents or human users. In both cases, retrieval is performed before language-model inference so that the generated response remains grounded in locally indexed mission and biological data.

The retrieval engine operates over the unified ChromaDB collection described in Subsection~\ref{subsubsec:embedding_indexing}. Because the collection contains both visual and textual records, the system supports four complementary retrieval modalities:
\begin{enumerate}
    \item text-to-text retrieval, in which a textual query is compared with indexed document chunks, annotations, and taxonomic prompts;
    \item text-to-image retrieval, in which a textual query retrieves semantically compatible reference images or mission observations;
    \item image-to-image retrieval, in which a query crop is compared with indexed visual reference embeddings;
    \item image-to-text retrieval, in which a query crop retrieves taxonomic prompts, species annotations, or other textual descriptions represented in the shared BioCLIP space.
\end{enumerate}

For a textual researcher query \(T_q\), the corresponding embedding is compared independently with the textual and visual partitions:
\begin{equation}
    \mathcal{R}_{t \rightarrow t}
    =
    \operatorname{TopK}
    \left(
        f_{\mathrm{text}}(T_q),
        \mathcal{C}_{\mathrm{text}}
    \right),
\end{equation}
\begin{equation}
    \mathcal{R}_{t \rightarrow i}
    =
    \operatorname{TopK}
    \left(
        f_{\mathrm{text}}(T_q),
        \mathcal{C}_{\mathrm{image}}
    \right),
\end{equation}
where \(\mathcal{C}_{\mathrm{text}}\) and \(\mathcal{C}_{\mathrm{image}}\) denote the textual and visual partitions of the unified collection. Similarly, when a query image or candidate crop \(I_q\) is available, the system performs image-to-image and image-to-text retrieval:
\begin{equation}
    \mathcal{R}_{i \rightarrow i}
    =
    \operatorname{TopK}
    \left(
        f_{\mathrm{img}}(I_q),
        \mathcal{C}_{\mathrm{image}}
    \right),
\end{equation}
\begin{equation}
    \mathcal{R}_{i \rightarrow t}
    =
    \operatorname{TopK}
    \left(
        f_{\mathrm{img}}(I_q),
        \mathcal{C}_{\mathrm{text}}
    \right).
\end{equation}
The compatibility of these four retrieval modes results from the shared multimodal embedding space learned by the CLIP-based encoder.

To prevent one modality from dominating the final context, the deployed retrieval configuration retains a controlled mixture of six visual records and four textual records for a standard top-\(10\) retrieval operation. The visual subset provides direct evidence from similar organisms, reference images, and mission frames, whereas the textual subset introduces explicit taxonomic annotations, scientific descriptions, sensor-log segments, and other structured information. This \(6{:}4\) composition controls the type of evidence provided to the reasoning agents.

The retrieval passes are executed separately and their results are interleaved before final selection. This prevents the substantially larger visual collection from occupying all available retrieval positions and ensures that textual evidence remains represented. When a modality filter is explicitly requested, retrieval can instead be restricted to either image or text records.

Each similarity-search pass initially retrieves more candidates than the final requested number. The candidates are ordered according to vector distance and then filtered to reduce redundancy. In particular, retrieved frames originating from the same video are subjected to temporal deduplication: frames occurring within a three-second interval of an already retained frame are discarded. This procedure reduces the repeated retrieval of nearly identical consecutive observations while preserving temporally distinct events.

Every retained result is represented as a LangChain \texttt{Document} containing its indexed content and associated metadata. The retrieval engine augments this metadata with the vector distance and the origin of the search pass. Results obtained from visual and textual queries are subsequently merged, duplicate identifiers are removed, and the final list is truncated to the configured retrieval size.

The metadata available to the reasoning layer depend on the retrieved modality and source dataset. Taxonomic reference records may provide
\texttt{species}, \texttt{genus}, \texttt{family}, \texttt{order}, and \texttt{class}, together with the source image path and file name. Mission video records may provide the source video, crop or frame path, acquisition timestamp, detection label, detection confidence, BioCLIP-2 prediction and similarity scores. Sensor-log and tabular records may include the source file, file name, and the start and end timestamps of the retrieved interval.

Before being passed to the QueryAgent or ReportAgent, the raw retrieval results are transformed into a structured context package:
\begin{equation}
    \mathcal{P}
    =
    \left\{
        C_{\mathrm{text}},
        \mathcal{I},
        \mathcal{S}_{\mathrm{image}},
        S_{\mathrm{retrieval}}
    \right\},
\end{equation}
where \(C_{\mathrm{text}}\) is the assembled textual context, \(\mathcal{I}\) is the set of loaded images, \(\mathcal{S}_{\mathrm{image}}\) contains their source paths, and \(S_{\mathrm{retrieval}}\) is a compact summary of the retrieved evidence.

Textual records are separated according to their origin. Sensor-log and CSV chunks are formatted with their source file and temporal interval:
\begin{quote}
\texttt{[Source: file | time: start \(\rightarrow\) end]}
\end{quote}
whereas biological annotation records are formatted using the available taxonomic hierarchy. For example, a fish annotation may contain the species, genus, family, order, and class followed by its descriptive text. Visual records also contribute textual metadata to the context. An identified reference image contributes its taxonomic annotation, whereas an unidentified mission frame contributes its source video or image file and acquisition timestamp.

The textual context is limited to a configurable maximum length, set to \(4000\) characters in the current implementation. When grouped video summaries are available, they are inserted before the remaining retrieved chunks so that the principal temporal and ecological evidence is preserved if truncation becomes necessary.

A maximum of four retrieved images is loaded into the multimodal context package. The images are accessed only after retrieval, resized to \(1344 \times 756\) pixels, converted to RGB, JPEG-encoded, and represented as base64 strings. This just-in-time image loading avoids placing complete image contents in the vector database while allowing the downstream reasoning model to inspect the most relevant visual evidence. The original file paths are retained separately for logging and traceability.

For video-oriented queries, the system additionally supports grouped crop retrieval. In this mode, image records belonging to a selected source video are first grouped using reliable taxonomic or detection metadata. A species label is considered when an authoritative species value is available, when the BioCLIP score exceeds the configured threshold, or when a sufficiently confident non-generic detection label is present. Records without reliable labels are grouped according to the similarity of their normalized visual embeddings.

Let \(\mathbf{g}_j\) denote the normalized centroid of an existing video group and \(\mathbf{v}_i\) the embedding of an unlabelled candidate crop. The candidate is assigned to the most similar group when
\begin{equation}
    \max_j
    \left(
        \mathbf{v}_i^{\top}\mathbf{g}_j
    \right)
    \geq
    \tau_g,
\end{equation}
where \(\tau_g\) is the configured grouping threshold. Otherwise, a new visual group is created. In the current configuration, the default similarity threshold is \(0.86\).

For each group, the system selects one representative crop using the strongest available metadata confidence and query similarity. The corresponding grouped summary contains the number of detections, first and last observation times, temporal segments, representative file, representative timestamp, best and mean confidence or similarity, and the available indexed-label evidence. When trained centroid or linear SVM models are available, the summary may also include the three strongest classification candidates at the class, order, family, genus, and species levels.

The grouped retrieval mode therefore prevents a long video sequence containing repeated observations of the same organism from overwhelming the reasoning context. Instead of returning every similar crop, it provides one visual representative per group together with a structured account of its occurrence frequency and temporal distribution.

The context packager finally produces a concise retrieval summary indicating the amount and type of evidence obtained, such as the number of sensor-log chunks, textual fish annotations, visual fish annotations, loaded images, and grouped video representatives. This summary allows the reasoning agents to determine whether sufficient evidence was retrieved before producing an ecological interpretation.

The resulting structured package is supplied to the local instruct model, Qwen3.5-4B, through the Ollama inference runner. The system employs this package to generate grounded answers for researchers and to produce detailed species reports, mission summaries, temporal activity descriptions, and anomaly explanations. Rather than relying exclusively on the parametric knowledge of the underlying language model, the framework cites retrieved source files, timestamps, taxonomic metadata, and visual observations directly.

In high-activity Mission Mode, the instruct model can remain resident in GPU memory to reduce repeated model-loading and generation latency. In autonomous or energy-constrained operating states, it can be unloaded after query execution to reduce steady-state memory use and power consumption.

Retrieval-augmented reasoning does not replace the species identification layer. The identification layer generates taxonomic hypotheses from the visual observations, whereas the retrieval layer gathers the biological, visual, temporal, and operational evidence associated with those hypotheses. The language model subsequently uses this evidence to explain and contextualize the observations without modifying the underlying taxonomic predictions.

\subsection{Multi-Agent Control Framework}
\label{subsec:multi_agent_control}

The high-level behavior of the monitoring station is controlled by a local multi-agent framework. This framework coordinates retrieval, structured analysis, energy management, reporting, hardware configuration, and autonomous post-processing. Instead of relying only on fixed rules, the system assigns specialized responsibilities to different agents. This makes the station more adaptable to changing mission objectives, environmental conditions, and power constraints.

\begin{figure*}
\centering
\includegraphics[width=1\linewidth]{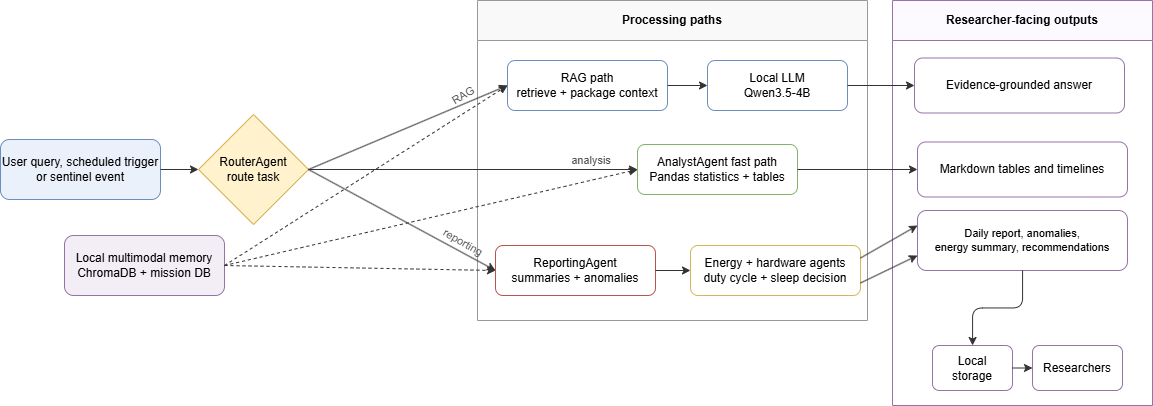}
\caption{Unified agentic reasoning and reporting layer. Incoming researcher queries or automated system triggers are evaluated by a RouterAgent, which selects the appropriate retrieval, analysis, or reporting pathway. The agents interact with local memory, the local instruct model, and hardware-management modules to generate grounded outputs for researchers.}
\label{fig:agentic_reasoning_layer}
\end{figure*}

\subsubsection{Router Agent}
\label{subsubsec:router_agent}

The Router Agent is responsible for orchestrating queries in a heterogeneous, multimodal search environment. Rather than relying on static mapping, it dynamically determines the target database collections, the search modalities (textual, visual, or hybrid), and the downstream processing pathway. 

To achieve this, the agent is structured as an interactive reasoning loop following the \textsc{Think--Act--Observe} paradigm. Before finalizing a retrieval route, the agent can actively probe candidate collections. Through a probing (or ``peeking'') mechanism, it retrieves and inspects small metadata samples from specific collections, using these empirical observations to resolve query ambiguity.

To assist the agent with complex taxonomic classification queries, the system incorporates a hybrid context-enrichment step before the reasoning loop begins. If a query contains taxonomic keywords or is accompanied by an image, the system runs a fast, non-parametric classification step using pre-calculated reference centroids. The top taxonomic predictions from this step are injected directly into the agent's reasoning context as high-confidence prior beliefs, guiding it toward the correct reference database.

For queries requesting visual species evidence (e.g., listing visually different species in a specific video), the agent enforces a crop-level grouping guardrail. This path bypasses naive global similarity retrieval and instead invokes a specialized visual-clustering pipeline that groups similar object detections. Alternatively, for quantitative or temporal queries (e.g., species occurrence timelines or video database inventories), the agent can activate an analytical fast path. This path bypasses the generative language model entirely, executing database-level metadata aggregations and reporting the results directly.

Finally, to handle edge cases or allow direct developer intervention, the architecture supports a deterministic target override. When an explicit collection target is provided by the user, the routing agent's reasoning loop is bypassed entirely, and queries are routed directly to the mapped reference or multimodal collections.

\subsubsection{Analyst Agent}
\label{subsubsec:analyst_agent}

The AnalystAgent is designed to resolve queries that are better addressed via structured computation than through heuristic generation. It compiles crop-level metadata (such as species predictions, BioCLIP-2 confidence scores, video names, and frame timestamps) extracted from the vector database into a local Pandas DataFrame. An auxiliary text-only model then generates targeted Python code to perform statistical tasks, including species occurrence counting, temporal distribution timelines, co-occurrence intervals, and detection-frequency estimations. We evaluated both Qwen3.5-4B and Ornith-9B; the latter demonstrated significantly superior performance, exceeding initial expectations.

To guarantee system safety and execution determinism, the generated code is executed within a restricted local namespace that isolates the system. File-system operations and hazardous python built-ins are dynamically blocked, and the code is statically validated using an abstract syntax tree (AST) parser prior to execution. If a runtime error is encountered, a bounded self-correction feedback loop is initiated to repair the code. The final outputs, including timestamps normalized to \texttt{HH:MM:SS} format, are printed and returned as structured Markdown tables or lists.

Because this computational pathway bypasses the larger vision-language instruct model for purely quantitative queries, it reduces system latency and power consumption.

\subsubsection{Energy and Sensor Management Agent}
\label{subsubsec:energy_sensor_management_agent}

The Energy and Sensor Management Agent is responsible for dynamically optimizing the power state of the station to ensure it survives the designated mission duration. Rather than relying on static power configurations, the agent continuously computes a real-time power budget by comparing remaining battery capacity ($\text{Wh}$) against the remaining mission time ($\text{hours}$). 

To evaluate current consumption, the agent monitors real-time system metrics—including CPU, GPU, and auxiliary power rails via Tegrastats on the primary Jetson compute module—alongside a constant baseline approximation of the always-on sentinel microcontrollers. Based on these variables and the recent event detection rate, the agent dynamically adjusts software- controlled parameters to align the power draw with the remaining energy reserves. These adjustments include tuning camera capture frame rates (FPS), lengthening or shortening autonomous compute cycle intervals, and modifying the keep-alive duration of cached model weights. 

Furthermore, the agent determines the sentinel confidence thresholds running on the microcontrollers to control hybrid wake-up triggers, deciding when to transition the primary compute module from a powered-off state directly into high-fidelity processing. In this manner, energy management is treated as an active, context-aware control loop that balances taxonomic accuracy with long-term hardware survivability on batteries.

\subsubsection{Dynamic Model Deployment Agent }
\label{subsubsec:dynamic_model_deployment_agent }

The Dynamic Model Deployment Agent manages the dynamic adaptation of the low-power sentinel tier. Ultra-low-power microcontrollers offer extreme energy efficiency but are constrained by strict on-chip memory limits, preventing the concurrent deployment of multi-task neural networks. The agent resolves this limitation by dynamically selecting and flashing specialized, highly-quantized neural network weights onto the microcontrollers based on deployment locations, real-time battery status, seasonal predictions, and researcher-defined mission priorities. 

For instance, the agent can reconfigure the sentinel layer to prioritize cetacean acoustic monitoring in one zone, and hot-swap to visual invasive species detection in another. To prevent flashing failures, the agent queries the model collection, verifying candidate model sizes, hardware register constraints, compatibility matrices, and past flashing histories before committing to a hardware update. This orchestration enables a highly specialized sentinel layer to maintain runtime flexibility across diverse long-term missions.

\subsubsection{Reporting Agent}
\label{subsubsec:reporting_agent}

The Reporting Agent converts the station's indexed video logs into researcher-facing situation summaries. Rather than presenting a flat list of event detections, the agent aggregates raw vector database metadata to produce a structured, high-level overview of the survey's progress and findings.

To generate a report, the agent compiles a deterministic database snapshot of the video assets. This snapshot includes global metadata coverage statistics, the number of processed video files, overall taxonomic species classification distributions, and chronologically segmented temporal detection events (grouping consecutive occurrences of individual species within a configurable temporal window). A local language model then synthesizes this raw structured data into a cohesive, scientist-oriented narrative.

The resulting Markdown (\texttt{.md}) report is saved to a designated local directory with a query-specific, timestamped filename. This workflow enables the low-bandwidth station to store and transmit compact, highly descriptive summaries rather than raw, high-resolution sensor streams when satellite or acoustic telemetry connections become available.

\section{Low-Power Sentinel Subsystems}
\label{sec:sentinel_subsystems}

This section details the two specialized low-consumption subsystems deployed around the Jetson master node: the visual detection subsystem on the MAX78002 and the acoustic detection subsystem on the MAX78000.

\subsection{Visual Detection Subsystem}
\label{subsec:visual_subsystem}

The proposed visual subsystem implements a sequential, shared-bus data architecture to optimize memory utilization during hardware-accelerated execution. Data acquisition begins through both a MIPI CSI-2 and DVP interface, which captures RGB565 image frames and streams them sequentially into external QSPI SRAM using a dedicated line handler. The system was tested with two cameras, OV5640 and OVM7692.

\begin{figure*}
    \centering
    \includegraphics[width=0.9\linewidth]{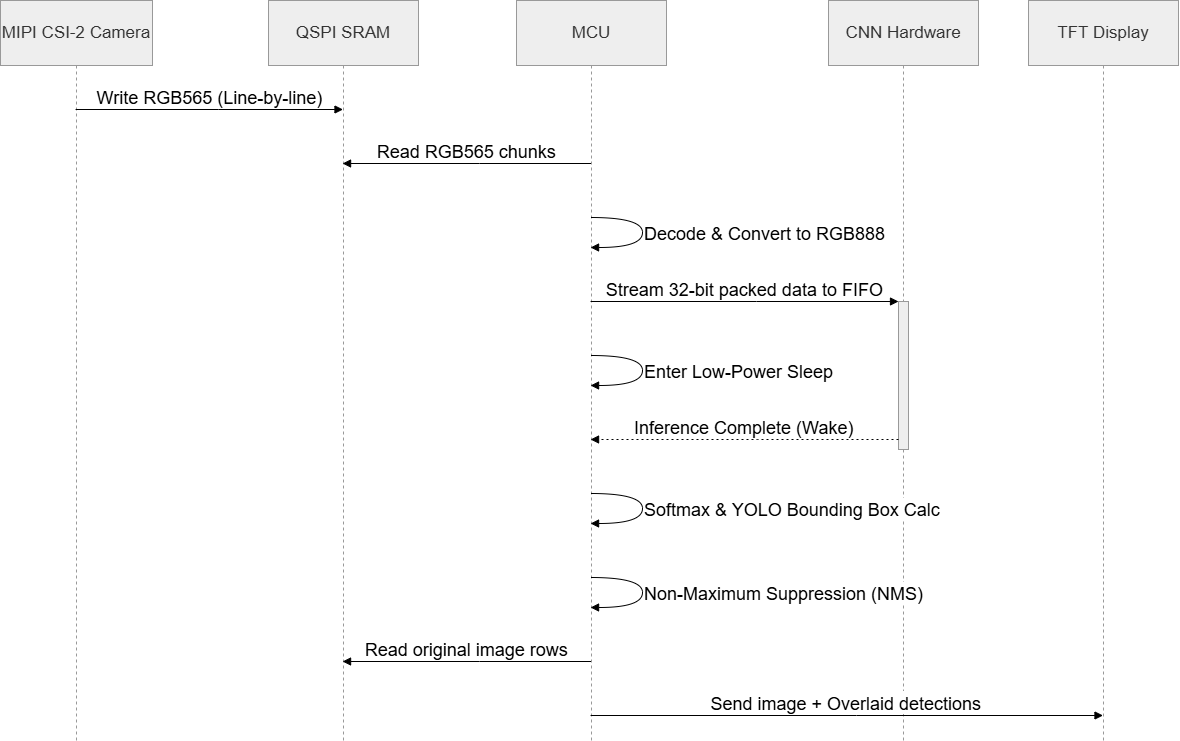}
    \caption{Visual detection subsystem.}
    \label{fig:visual_subsystem}
\end{figure*}

During preprocessing, the microcontroller unit (MCU) retrieves the stored frame in distinct chunks, transcodes the pixel data into 24-bit RGB888 format, packs the results into 32-bit words, and loads them directly into the convolutional neural network (CNN) hardware FIFO. To minimize overall energy consumption, the MCU transitions into a low-power sleep state during inference and delegates computation of bounding box regressions and class logits to the hardware accelerator. Upon inference completion, the MCU resumes execution to poll the output registers. It applies a softmax activation to derive class probabilities and transforms the spatial outputs into YOLO-style bounding box coordinates. Redundant localizations are systematically eliminated using a Non-Maximum Suppression (NMS) algorithm based on Intersection over Union (IoU) filtering. Finally, the raw image is retrieved row by row from SRAM and transmitted over SPI to a TFT display with overlaid bounding boxes and confidence scores. While real-time display rendering is primary for validation, operational deployments bypass the display, opting instead to write the image to an SD card or issue a GPIO interrupt to wake the Jetson host module. Figure \ref{fig:visual_subsystem} represents the sequence diagram of this subsystem. 

Two lightweight vision models are tested: FPN Detector and TinyissimoYOLO. Their architectures are described in Sections~\ref{subsubsec:fpn_detector} and~\ref{subsubsec:tinyissimoyolo}. A major constraint on the MAX78002 is that the hardware dictates how the model should be structured. The architecture supports up to 128 layers, 2048 input and output channels per layer, and image dimensions up to 2047 pixels. Supported operations include Conv1D, Conv2D, ConvTranspose2D, pooling, fully connected layers, and element-wise arithmetic operations. Weights can be quantized to 1, 2, 4, or 8 bits, and activations are processed in signed 8-bit format.

The device provides approximately 2.34 MiB of dedicated weight memory and 1.28 MiB of data SRAM distributed across 16 memory banks. Convolutional layers are restricted to specific kernel sizes and strides optimized for hardware efficiency, with support for depthwise convolutions and streaming mode to process inputs exceeding on-chip memory capacity. Convolutional layers on the MAX78002 are limited to the following hardware-supported configurations:
\begin{itemize}
    \item \textbf{Conv2D:} kernel size $1\times1$ or $3\times3$ only.
    \item \textbf{Conv2D stride:} fixed to $1\times1$.
    \item \textbf{Conv2D padding:} $0$, $1$, or $2$.
    \item \textbf{Conv2D dilation:} $1$ to $16$.
    \item \textbf{Groups:} standard convolution ($g=1$) or depthwise convolution ($g=C_{\text{in}}=C_{\text{out}}$).
    \item \textbf{Conv1D:} kernel size $1$ to $9$.
    \item \textbf{Conv1D stride:} fixed to $1$.
    \item \textbf{Conv1D padding:} $0$ to $2$.
    \item \textbf{ConvTranspose2D:} kernel size $3\times3$ only.
    \item \textbf{ConvTranspose2D stride:} fixed to $2\times2$.
    \item \textbf{ConvTranspose2D output padding:} fixed to $1$.
\end{itemize}

Consequently, convolutions such as $5\times5$, $7\times7$, stride-$2$ Conv2D, stride-$4$ Conv2D, or arbitrary grouped convolutions are not directly supported by the accelerator and must be decomposed into supported operations during model conversion. Fully connected layers are implemented as $1\times1$ convolutions and are constrained by input flattening limits. The platform also uses hardware-accelerated activation functions.

\subsubsection{FPN Detector}
\label{subsubsec:fpn_detector}

The first tested model is a Feature Pyramid Network (FPN) object detection architecture developed by the MAX 78002 team and tailored specifically for ultra-low-power, real-time edge AI processing. Operating on an RGB input image of size \(256 \times 320\), the network uses two initial fused convolutional layers to project the input to a 64-channel space before feeding it into a highly optimized \texttt{ResNetBackbone}. This backbone uses strided max-pooling layers directly embedded within its residual blocks to downsample the spatial dimensions across six distinct sub-blocks, producing four multiscale feature maps at resolutions matching \(32\times40\), \(16\times20\), \(8\times10\), and \(4\times5\).

These representations are then aggregated by a top-down FPN module that uses \(1\times1\) lateral convolutions to unify all pyramid feature planes to a 64-channel baseline, together with \(3\times3\) transposed convolutions for spatial upsampling and element-wise additions to reduce aliasing artifacts. The resulting pyramid feature maps, \(\mathbf{P}_0\) to \(\mathbf{P}_3\), are routed in parallel to dual, weight-sharing subnetworks for classification and bounding box regression. These subnetworks simultaneously evaluate 10,200 anchor boxes per frame generated across variable aspect ratios \((0.5, 1.0, 2.0)\) and scale modifiers.

By using proprietary fused primitives the architecture maximizes hardware pipelining efficiency. This allows complex structural semantics and anchor-based post-processing operations, including Jaccard-coordinate decoding and Non-Maximum Suppression, to execute within the strict energy and memory constraints of the MAX78002 microcontroller platform. More details are provided in the repository and documentation supplied by Analog Devices \citep{AnalogDevices2022MAX78002}.

\subsubsection{TinyissimoYOLO}
\label{subsubsec:tinyissimoyolo}

TinyissimoYOLO is an ultra-lightweight object detection architecture engineered specifically for extreme edge-computing environments and resource-constrained microcontrollers. It was developed by Julian Moosmann at the Center for Project Based Learning, ETH Zurich \citep{Moosmann2023TinyissimoYOLO}. The network relies on a streamlined, strictly sequential convolutional neural network backbone consisting of standard 2D convolutional layers, typically using \(3\times3\) kernels with ReLU activations, interspersed with \(2\times2\) max-pooling layers.

This topology progressively downsamples the spatial dimensions of the input image while expanding channel depth, generally scaling from 16 up to 128 channels. To strictly minimize computational overhead and memory footprint, the architecture avoids complex mechanisms such as feature pyramids. Instead, the final pooled feature map is routed directly into a compact detection head, often implemented as a dense layer or a \(1\times1\) convolution, which outputs a structured tensor containing bounding box coordinates and class probabilities. This design prioritizes low parameter count and deterministic execution, enabling real-time inference on the MAX78002 microcontroller. Additional implementation details and model variants are provided in the associated repository \citep{Moosmann2023TinyissimoYOLO}.

\subsection{Acoustic Detection Subsystem}
\label{subsec:acoustic_subsystem}

The developed acoustic classification system implements a direct-memory-access-driven continuous processing pipeline to detect marine mammal vocalizations on a resource-constrained hardware architecture. Audio is sampled at 16 kHz through an I2S interface and buffered dynamically. Incoming data is digitally high-pass filtered to eliminate DC offset and evaluated against an amplitude threshold to suppress inference on ambient noise.

For valid acoustic events, a software-based digital signal processing routine generates a log-Mel spectrogram by applying a periodic Hann window, executing a 512-point real Fast Fourier Transform, and mapping the power spectrum through a 64-bin sparse triangular Mel filterbank. The resulting feature matrix is quantized to 8-bit integers and loaded directly into the static random-access memory of a dedicated convolutional neural network accelerator.

The deployed classifier employs a compact five-stage convolutional neural network optimized for execution on the MAX78000 hardware accelerator. The network accepts a single-channel 96×64 log-Mel spectrogram as input and progressively extracts higher-level acoustic features through successive 3×3 convolutional layers, each followed by batch normalization, ReLU activation, and 2×2 max-pooling (except for the first layer). The feature maps are reduced from 16 to 64 channels while the spatial resolution is gradually compressed, yielding a compact 64×3×2 representation. A final 1×1 convolution preserves the feature depth before flattening the output into a 384-element feature vector, which is processed by a fully connected layer to classify the input into one of 13 categories, comprising 12 cetacean vocalization classes and one background noise class. The architecture is specifically designed to minimize computational complexity and memory usage while maintaining sufficient representational capacity for accurate embedded acoustic classification.

\begin{figure*}
    \centering
    \includegraphics[width=0.8\linewidth]{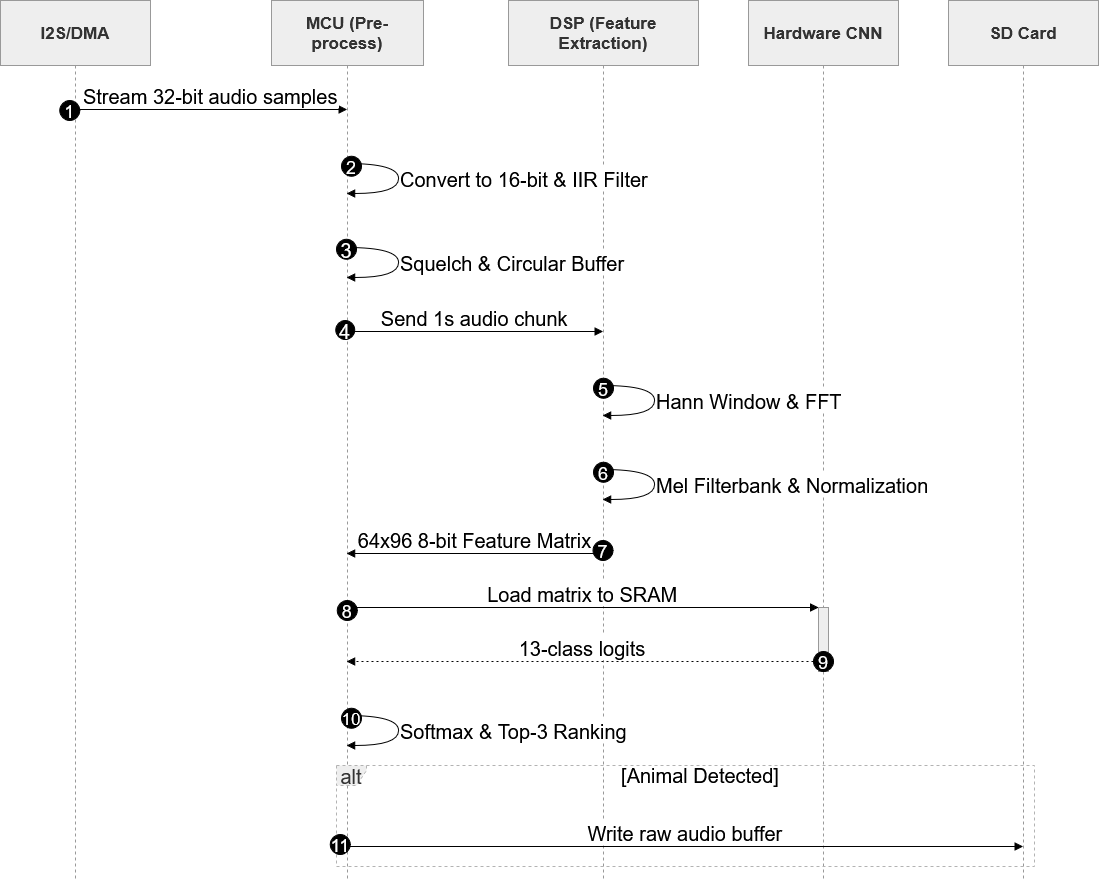}
    \caption{Acoustic detection subsystem.}
    \label{fig:acoustic_subsystem}
\end{figure*}

During CNN inference, the primary microcontroller yields execution to the hardware accelerator to reduce dynamic power consumption. Once inference is complete, the accelerator returns a vector of class scores, which is processed using a softmax function to obtain the confidence associated with each category. If the highest-confidence prediction corresponds to a cetacean vocalization rather than the background noise class, the associated raw audio buffer is automatically archived to the integrated SD card for subsequent empirical validation. The complete processing sequence of the acoustic subsystem is illustrated in Figure \ref{fig:acoustic_subsystem}.

\section{Experimental Results and Evaluation}
\label{sec:results}

This section evaluates the operational efficiency, detection accuracy, and system-wide energy consumption of the proposed hierarchical agentic underwater monitoring system. The evaluation is structured to first benchmark the ultra-low-power sentinel tier independently, and then to validate the high-fidelity master tier and the overall power savings achieved by the hardware-switched cascade architecture.

\subsection{Experimental Setup and Benchmarking Environment}
\label{subsec:experimental_setup}

To validate the system under realistic edge constraints, the multi-agent framework and high-fidelity models were executed on a Seeed reComputer J4012 powered by an NVIDIA Jetson Orin NX module (16GB RAM) . The continuous sentinel tier consisted of a MAX78002 EV kit and a MAX78000 FTHR board. The NVIDIA Jetson Orin NX and Maxim Integrated MAX78000/MAX78002 were powered by an Aim-TTi CPX400DP DC power supply. System current, voltage, and total power consumption for the MAX78000 FTHR board and the Seeed reComputer J4012 (Jetson Orin NX) were monitored and recorded separately using an external, inline Rohde \& Schwarz HMC8015 digital power analyzer. In contrast, the MAX78002 EV Kit tracked its power consumption internally using its onboard power accumulator module.

The evaluation datasets were already introduced in section \ref{sec:datasets}.

\subsection{Sentinel Evaluation}

The Sentinel Tier was evaluated under the deployment constraints targeted by the proposed low-power architecture. The visual sentinel was assessed using mAP@50, while the acoustic sentinel was evaluated using accuracy, recall and F1 score. The visual sentinel was trained for 100 epochs using the same train/validation/test split proposed by the original dataset authors. Training employed the Adam optimizer with an initial learning rate of 0.001, a batch size of 64, and a cosine annealing learning rate schedule. Quantization-aware training (QAT) was enabled from epoch 73 onward. As described in Section~\ref{subsec:fishdetm}, stochastic data augmentation was applied throughout training, which was further restricted to images containing at most five annotated bounding boxes to better reflect the target deployment scenario. All visual model training was performed on the IFREMER HPC cluster. For the acoustic sentinel, evaluation was conducted using five-fold cross-validation. The model was trained for 50 epochs using the Adam optimizer (weight decay $10^{-4}$), an initial learning rate of 0.001, and a batch size of 128, with QAT enabled from epoch 37 onward. Table~\ref{tab:system_comprehensive_metrics} summarizes the resulting performance of both Sentinel Tier models.

\begin{table*}[htbp]
\centering
\caption{Sentinel-tier core performance evaluation metrics.}
\label{tab:system_comprehensive_metrics}
\scriptsize
\begin{tabular}{llcc}
\toprule
\textbf{Subsystem (Chip)} & \textbf{Model Architecture} & \textbf{Target Task} & \textbf{Evaluation Metrics} \\
\midrule
\multirow{2}{*}{Vision (MAX78002)} & FPN Detector & Fish Detection & mAP@50: 78.4\% \\
 & TinyissimoYOLO & Fish Detection & mAP@50: 70.75\% \\
\midrule
Acoustic (MAX78000) & Mel-Spectrogram CNN & 12 Marine Mammals & \shortstack[l]{Accuracy: 98.99 $\pm$ 0.11\% \\ Recall: 95.97 $\pm$ 0.86\% \\ F1: 96.36 $\pm$ 0.52\%} \\
\bottomrule
\end{tabular}
\end{table*}

\subsection{Experimental Validation of the Master Tier}
\label{subsec:master_tier_evaluation}

The primary objective of the experimental evaluation was to verify the feasibility of the proposed edge-based multi-agent system under realistic deployment constraints. While the fish identification component could be quantitatively assessed using curated benchmark datasets, evaluating the complete agentic RAG pipeline proved considerably more challenging due to the absence of standardized datasets and established metrics tailored to this application. Furthermore, the scope of this work did not permit the development of dedicated benchmarks for the routing, analyst, and reporting agents. Consequently, the evaluation combines rigorous quantitative experiments for the fish identification module with qualitative validation of the remaining agents, focusing on their functional correctness, integration, and suitability for low-power edge deployment.

The fish identification layer was evaluated using two complementary protocols. First, a cross-dataset evaluation was conducted across the six reference collections described in Table~\ref{tab:datasets}: FishBase/SeaLifeBase, WildFish++, Fish-Vista, FishNet, BioTrove restricted to marine species, and OzFish. In this setting, each dataset was used as a reference collection and evaluated against the others in order to quantify cross-domain generalization. OzFish was used as the real-condition reference dataset, making it the main target for assessing deployment robustness. Since the retrieval and classification pipeline is deterministic once the embeddings are fixed, this cross-dataset evaluation was performed once.

Second, an intra-dataset evaluation was conducted independently for each dataset. For this protocol, 10 independent species-stratified splits were generated, with 80\% of the samples embedded into the vectorized reference space and the remaining 20\% used for testing. Results are reported as mean $\pm$ standard deviation. Classification was evaluated at five taxonomic levels: class, order, family, genus, and species. Two ranking-based metrics were used: Top-1 accuracy and Top-5 accuracy. The cross-dataset results are summarized in Table~\ref{tab:cross_dataset_matrix}, while the intra-dataset split results are reported in Table~\ref{tab:identification_evaluation}.

For both cross-dataset and intra-dataset evaluations, the tables present results for the SVM classifier, which achieved superior performance in most cases. This outcome was expected, as centroid-based methods risk obscuring intra-species fine details, particularly when morphology and coloration vary across age or sex.

The remaining components of the system were evaluated qualitatively through functional validation. Given the specialized nature of the proposed architecture, no publicly available benchmarks currently exist to objectively assess the performance of an edge-deployed multi-agent RAG system operating across heterogeneous scientific databases. Consequently, the evaluation focused on verifying that the agents performed their intended roles reliably while satisfying the power and computational constraints required for autonomous edge deployment.

For the routing architecture, testing confirmed the correct selection of the appropriate knowledge source and successful execution of the retrieval pipeline under realistic operating conditions. Although large-scale benchmarking involving numerous vector databases with overlapping semantic domains constitutes an important direction for future work, the present experiments demonstrate the practical feasibility of the routing strategy within the target hardware and energy constraints.

Testing of the AnalystAgent confirmed that structured code generation and execution can be performed directly on the Jetson edge platform. In initial evaluations, the Qwen3.5-4B model exhibited significant performance limitations on complex tasks: the ReAct execution loop frequently required up to eight iterations to reach a solution, primarily due to constraints on available functions/libraries and an inability to process the full context simultaneously. In contrast, testing with the Ornith 9B model exceeded expectations, generating executable Pandas code on the first iteration in most cases, and requiring at most two loops. Furthermore, because Ornith 9B is a text-only model, it eliminates the unnecessary computational overhead of vision-language modalities present in Qwen3.5-4B—a key advantage for the text- and data-centric requirements of the AnalystAgent. One operational requirement identified on the Jetson platform is the necessity of clearing the NVIDIA GPU cache prior to initializing the AnalystAgent to prevent CUDA out-of-memory (cudaMalloc) errors. Overall, these results demonstrate that deploying capable 8B–9B parameter models on resource-constrained edge hardware achieves both high execution accuracy and system feasibility.

The ReportingAgent was not evaluated using a dedicated quantitative metric. Its behavior is largely deterministic, producing structured Markdown reports from the validated outputs of the upstream agents and the current station state. Consequently, its assessment focused on verifying the correctness, consistency, and completeness of the generated reports rather than measuring predictive performance.

\newcommand{\taxocell}[5]{%
\begin{tabular}{@{}c@{}}
\scriptsize C: #1\\
\scriptsize O: #2\\
\scriptsize F: #3\\
\scriptsize G: #4\\
\scriptsize S: #5
\end{tabular}}

\begin{table*}[htbp]
\centering
\caption{Cross-dataset taxonomic identification matrix. Rows indicate the dataset used as the embedded reference collection, while columns indicate the dataset used for testing. Each cell reports a compact taxonomic performance signature in the order Class, Order, Family, Genus, and Species. Values are reported as Top-1/Top-5 accuracy percentages. OzFish is used as the real-condition deployment reference.}
\label{tab:cross_dataset_matrix}
\scriptsize
\setlength{\tabcolsep}{2.5pt}
\renewcommand{\arraystretch}{1.15}
\begin{tabular}{lcccccc}
\hline
\textbf{Reference $\rightarrow$ Test}
& \textbf{FishBase/} 
& \textbf{WildFish++}
& \textbf{Fish-Vista}
& \textbf{FishNet}
& \textbf{BioTrove}
& \textbf{OzFish} \\
& \textbf{SeaLifeBase}
&
&
&
& \textbf{Marine}
& \textbf{Real} \\ \hline

\textbf{FishBase/SeaLifeBase}
& --
& \taxocell{98.7/100.0}{78.5/95.1}{75.4/90.4}{65.3/80.0}{51.9/67.8}
& \taxocell{99.3/100.0}{81.5/96.7}{75.4/91.1}{45.3/66.0}{11.2/23.9}
& \taxocell{98.0/99.9}{80.6/96.6}{73.3/88.8}{70.4/84.4}{60.0/70.6}
& \taxocell{97.7/99.8}{75.3/95.9}{60.6/83.6}{48.2/71.6}{24.7/43.1}
& \taxocell{97.4/100.0}{73.0/92.9}{13.4/42.9}{13.9/30.0}{3.2/6.8} \\

\textbf{WildFish++}
& \taxocell{88.5/89.7}{66.4/83.3}{46.3/57.5}{35.1/41.5}{23.5/24.7}
& --
& \taxocell{99.4/100.0}{70.3/94.5}{67.2/83.5}{21.7/32.6}{8.4/19.1}
& \taxocell{98.4/99.9}{77.9/94.6}{66.6/79.4}{52.3/61.4}{39.4/44.8}
& \taxocell{98.3/99.6}{71.6/93.0}{56.4/72.4}{37.1/49.3}{20.5/26.1}
& \taxocell{95.5/100.0}{70.1/93.4}{15.3/44.5}{11.5/27.7}{2.6/6.7} \\

\textbf{Fish-Vista}
& \taxocell{84.8/85.3}{53.3/71.7}{25.9/38.7}{12.0/18.5}{3.4/4.7}
& \taxocell{96.2/96.5}{62.2/83.6}{40.2/61.1}{19.7/28.3}{5.4/7.5}
& --
& \taxocell{92.5/93.7}{59.7/80.8}{36.3/52.9}{19.2/28.1}{9.2/14.6}
& \taxocell{93.0/93.6}{57.4/78.1}{34.8/55.3}{11.5/20.3}{1.9/3.6}
& \taxocell{98.6/98.6}{75.6/89.6}{16.2/52.5}{13.2/29.0}{0.7/2.1} \\

\textbf{FishNet}
& \taxocell{88.8/90.0}{72.2/87.5}{57.2/68.3}{64.2/70.5}{65.0/67.2}
& \taxocell{98.9/100.0}{77.1/96.2}{73.2/90.4}{64.3/81.0}{47.2/65.9}
& \taxocell{99.3/99.9}{79.0/96.1}{74.9/91.1}{38.5/62.4}{14.9/30.6}
& --
& \taxocell{98.4/99.9}{73.7/95.5}{63.0/82.0}{41.3/65.0}{17.1/33.6}
& \taxocell{95.5/100.0}{70.4/93.3}{14.1/33.9}{9.6/19.8}{1.2/2.8} \\

\textbf{BioTrove Marine}
& \taxocell{88.5/89.6}{68.1/85.8}{46.7/63.1}{35.9/52.7}{19.7/32.0}
& \taxocell{98.9/99.8}{77.2/95.5}{72.6/88.9}{58.8/74.4}{44.3/59.9}
& \taxocell{99.1/99.7}{74.6/94.5}{65.9/85.2}{34.3/58.1}{3.0/8.5}
& \taxocell{98.4/99.8}{79.0/96.2}{69.0/85.5}{58.0/75.9}{41.1/56.8}
& --
& \taxocell{96.2/100.0}{73.5/94.1}{16.7/47.5}{12.6/33.4}{4.2/8.2} \\

\textbf{OzFish Real}
& --
& --
& --
& --
& --
& -- \\ \hline
\end{tabular}

\vspace{0.5em}
\footnotesize
Each non-diagonal cell reports five rows: C, O, F, G, and S for Class, Order, Family, Genus, and Species. Each row follows the format Top-1/Top-5 accuracy.
\end{table*}

\begin{table*}[htbp]
\centering
\caption{Intra-dataset taxonomic identification evaluation across validated marine reference datasets. Results are reported as mean $\pm$ standard deviation over 10 species-stratified 80/20 splits.}
\label{tab:identification_evaluation}
\small
\setlength{\tabcolsep}{2.8pt}
\renewcommand{\arraystretch}{1.15}
\begin{tabular}{lcccccccccc}
\hline
\multirow{2}{*}{\textbf{Dataset}} 
& \multicolumn{2}{c}{\textbf{Class}}
& \multicolumn{2}{c}{\textbf{Order}}
& \multicolumn{2}{c}{\textbf{Family}}
& \multicolumn{2}{c}{\textbf{Genus}}
& \multicolumn{2}{c}{\textbf{Species}} \\
\cline{2-11}
& \textbf{T1} & \textbf{T5}
& \textbf{T1} & \textbf{T5}
& \textbf{T1} & \textbf{T5}
& \textbf{T1} & \textbf{T5}
& \textbf{T1} & \textbf{T5} \\ \hline

Fish/SeaLife Base 
& 96.8 $\pm$ 0.1 & 99.6 $\pm$ 0.0
& 79.4 $\pm$ 0.2 & 96.9 $\pm$ 0.1
& 70.4 $\pm$ 0.2 & 88.6 $\pm$ 0.2
& 63.1 $\pm$ 0.2 & 83.4 $\pm$ 0.1
& 43.0 $\pm$ 0.4 & 61.3 $\pm$ 0.3 \\

WildFish++ 
& 99.6 $\pm$ 0.0 & 100.0 $\pm$ 0.0
& 86.3 $\pm$ 0.2 & 99.4 $\pm$ 0.1
& 90.3 $\pm$ 0.1 & 98.8 $\pm$ 0.0
& 88.7 $\pm$ 0.2 & 97.7 $\pm$ 0.1
& 79.4 $\pm$ 0.2 & 92.5 $\pm$ 0.1 \\

Fish-Vista 
& 99.7 $\pm$ 0.0 & 100.0 $\pm$ 0.0
& 95.3 $\pm$ 0.2 & 99.6 $\pm$ 0.1
& 95.4 $\pm$ 0.1 & 98.9 $\pm$ 0.1
& 83.9 $\pm$ 0.3 & 93.2 $\pm$ 0.1
& 68.9 $\pm$ 0.3 & 85.7 $\pm$ 0.2 \\

FishNet 
& 98.9 $\pm$ 0.0 & 100.0 $\pm$ 0.0
& 86.3 $\pm$ 0.1 & 98.5 $\pm$ 0.0
& 78.9 $\pm$ 0.1 & 90.6 $\pm$ 0.1
& 75.8 $\pm$ 0.2 & 89.4 $\pm$ 0.1
& 60.2 $\pm$ 0.2 & 73.3 $\pm$ 0.1 \\

BioTrove Marine 
& 98.7 $\pm$ 0.0 & 100.0 $\pm$ 0.0
& 80.8 $\pm$ 0.2 & 98.2 $\pm$ 0.1
& 69.6 $\pm$ 1.1 & 86.0 $\pm$ 0.7
& 73.8 $\pm$ 0.4 & 91.1 $\pm$ 0.2
& 68.9 $\pm$ 0.3 & 85.4 $\pm$ 0.1 \\

\hline

\end{tabular}

\vspace{0.5em}
\footnotesize
T1 and T5 denote Top-1 and Top-5 accuracy, respectively. Each value is reported as mean $\pm$ standard deviation across 10 independent species-stratified splits.
\end{table*}

\subsubsection{Energy Consumption Evaluation}
\label{subsubsec:energy_consumption_evaluation}

\paragraph{Continuous sentinel-layer consumption.}

The visual and acoustic sentinels remain active continuously, establishing the baseline energy footprint of the system. This continuous sentinel layer operates independently of the downstream processing cycles, and its energy contribution is measured in isolation to define the static power draw required for uninterrupted environmental monitoring. Table~\ref{tab:visual_sentinel_consumption} presents the experimental evaluation of the visual sentinel subsystem. We report both performances of Tinyissimo YOLO and FPN Detector. The evaluation considers eight distinct configurations: for both tested models employing either an OV5640 (a 5 MP CSI camera) or an OVM7692 (a VGA DVP camera), with measurements taken for both inference-only and capture-plus-inference operational modes. Notably, the OVM7692 yields significantly lower power consumption than the OV5640. This improvement is primarily attributed to its native hardware optimization for the MAX78002 EV Kit—with which it is bundled as the reference sensor—as well as highly tunable operational parameters, particularly regarding clock speed control. Furthermore, the OVM7692 benefits from reduced sensor startup latency, lower initialization overhead, and a simpler interface clocking scheme inherent to its low-resolution DVP architecture compared to the higher-resolution MIPI-CSI setup.

\begin{table*}[t]
\centering
\caption{Experimental characterization of the continuously active visual sentinel subsystem (MAX78002). Note: Values reflect only the deployable portion of the system, measured via the MAX78002 EVKIT internal power monitoring circuitry, excluding the rest of the development board components.}
\label{tab:visual_sentinel_consumption}
\scriptsize
\begin{tabular}{llccccc}
\toprule
Model & Camera \& Mode & 
\begin{tabular}[c]{@{}c@{}}Energy Per Inference\\ (mJ/inf)\end{tabular} & 
\begin{tabular}[c]{@{}c@{}}Active Power\\ mW\end{tabular} &
\begin{tabular}[c]{@{}c@{}}Idle Power\\ mW\end{tabular} &
\begin{tabular}[c]{@{}c@{}}Latency\\ ms\end{tabular} &
\begin{tabular}[c]{@{}c@{}}FPS\\ ms\end{tabular} \\

\midrule
\multirow{4}{*}{TinyissimoYOLO} 
 & OVM7692 (Inference Only)    & 9.2 & 90.65 & 11.3 & 116.4 & 8.6 \\
 & OVM7692 (Capture + Inf)     & 11.97 & 86.57 & 11.3 & 159 & 6.3 \\
 & OV5640 (Inference Only)     & 17.97 & 119 & 13.2 & 170 & 5.88 \\
 & OV5640 (Capture + Inf)      & 23.65 & 103 & 13.57 & 264.3 & 3.8 \\
\midrule
\multirow{4}{*}{FPN Detector} 
 & OVM7692 (Inference Only)    & 63.9 & 482.85 & 11.5  & 135 & 7.4 \\
 & OVM7692 (Capture + Inf)     & 67.13 & 372.56 & 11.31 & 185.8 & 5.4\\
 & OV5640 (Inference Only)     & 63.5 & 480 & 13.2 & 135 & 7.4 \\
 & OV5640 (Capture + Inf)      & 68.37 & 336.8 & 13.21 & 211.3 &  4.7 \\
\bottomrule
\end{tabular}
\end{table*}

\begin{table*}[t]
\centering
\caption{Experimental characterization of the continuously active audio sentinel subsystem (MAX78000FTHR). Note: Due to hardware monitoring limitations, values represent total board consumption, as isolated measurement of audio acquisition, CNN, and DSP components only is not supported.}
\label{tab:audio_sentinel_consumption}
\scriptsize
\begin{tabular}{lc}
\toprule
Audio Mode & 
\begin{tabular}[c]{@{}c@{}}Average Power\\ (mW)\end{tabular} \\
\midrule
Internal Microphone            & \(50\) \\
External Audio (FTHR CODEC)    & \(180\)  \\
\bottomrule
\end{tabular}
\end{table*}

\paragraph{Measured processing blocks.}

In contrast to the continuous sentinel layer, the Jetson module's energy consumption is highly dynamic and depends strictly on the deployment configuration. The system utilizes three distinct operating modes: periodic duty-cycling, event-triggered activation upon sentinel detection, and sustained continuous operation featuring a graphical user interface. 

To evaluate the energy requirements across these varying modes, the Jetson processing pipeline is decomposed into nine discrete, experimentally measurable operational blocks. These blocks, characterized in Table~\ref{tab:measured_processing_blocks}, correspond to the sequential tasks executed during system operation:

\begin{enumerate}
    \item boot and initialization;
    \item sentinel-data transfer;
    \item video-frame extraction;
    \item CLIP \& Grounding Dino loading;
    \item image/frame indexing;
    \item CLIP similarity retrieval;
    \item Router--RAG request processing;
    \item scientific report generation;
    \item controlled shutdown.
\end{enumerate}

\begin{table*}[t]
\centering
\caption{Experimental energy characterization of the discrete processing blocks utilized across the three deployment modes. Values are reported for the exact workload specified during each measurement.}
\label{tab:measured_processing_blocks}
\scriptsize
\begin{tabular}{lllll}
\toprule
Processing block &
Measured workload &
Energy (\(\mathrm{Wh}\)) &
Duration (\(\mathrm{s}\)) &
Average power (\(\mathrm{W}\)) \\
\midrule
Boot and initialization
& One activation
& \(0.0434\)
& \(16.374\)
& \(9.55\) \\

Sentinel-data transfer
& \(1~\mathrm{MB}\) (32 audio files of 32KB)
& \(0.26\)
& \(136\)
& \(6.88\) \\

Video-frame extraction
& \(1\) frame \(720*360\)
& \(0.0001\)
& \(0.038\)
& \(9.77\) \\

CLIP \& Grounding Dino loading
& One loading sequence
& \(0.076\)
& \(32\)
& \(8.55\) \\

Image/frame indexing
& \(1\) frame \(720*360\)
& $4.17 \times 10^{-5}$
& \(0.0137\)
& \(10.96\) \\

CLIP similarity retrieval
& One query over \(49150\) indexed items
& \(0.046\)
& \(20\)
& \(8.28\) \\

Router--RAG request
& One inference request
& \(0.431\)
& \(139\)
& \(11.16\) \\

Scientific report generation
& One report from 187 indexed detections
& 0.646
& 177
& 13.14 \\

Controlled shutdown
& One shutdown sequence
& \(0.014\)
& \(8\)
& \(6.3\) \\
\bottomrule
\end{tabular}
\end{table*}

\paragraph{Mathematical energy formulation.}

To extrapolate the total daily energy consumption (\(E_{\mathrm{total}}\)) across the three deployment configurations, the system's energy footprint is modeled as the sum of the static sentinel baseline (\(E_{\mathrm{sentinel}}\)) and the dynamic Jetson processing load (\(E_{\mathrm{dyn}}\)):
\begin{equation}
    E_{\mathrm{total}} = E_{\mathrm{sentinel}} + E_{\mathrm{dyn}}
\end{equation}

The continuous sentinel baseline is deterministic and independent of the operating mode. It is defined as:
\begin{equation}
    E_{\mathrm{sentinel}} = P_{\mathrm{sentinel}} \cdot T_{\mathrm{day}}
\end{equation}
where \(P_{\mathrm{sentinel}}\) is the combined average power of the active sentinels (Tables~\ref{tab:audio_sentinel_consumption} and ~\ref{tab:visual_sentinel_consumption}) and \(T_{\mathrm{day}}\) is the 24-hour operational period.

The dynamic energy, \(E_{\mathrm{dyn}}\), represents the variable consumption of the Jetson module. Let \(E_{b_i}\) denote the energy required to execute the \(i\)-th processing block (Table~\ref{tab:measured_processing_blocks}), and \(n_i\) denote its daily execution frequency. Because the individual energy blocks already incorporate the idle power draw and the time between operations is negligible, the dynamic load is formulated solely as the sum of these discrete processing events:
\begin{equation}
    E_{\mathrm{dyn}} = \sum_{i=1}^{9} n_i \cdot E_{b_i}
\end{equation}

The execution frequency \(n_i\) strictly depends on the selected operating mode:
\begin{itemize}
    \item \textbf{Periodic duty-cycling:} The Jetson powers on at a fixed time interval \(\Delta t\). The execution frequency for core retrieval and indexing blocks is deterministic, defined as \(n_i = \lfloor T_{\mathrm{day}} / \Delta t \rfloor\). The system undergoes a controlled shutdown (Block 9) after each cycle.
    \item \textbf{Event-triggered:} The processing sequence is initiated exclusively by sentinel detection events. The frequency \(n_i\) becomes a stochastic variable corresponding to the expected daily event rate, \(\lambda_{\mathrm{event}}\).
    \item \textbf{Sustained continuous:} The Jetson module remains powered indefinitely to support the graphical interface and real-time processing. The execution frequency \(n_i\) is driven by the volume of spontaneous scientist requests (Block 7).
\end{itemize}

By defining hypothetical operational scenarios to establish the frequency variables (\(n_i\)), this formulation, combined with the measured energy blocks, provides a direct method to estimate the total daily energy consumption of the system. This consumption profile is subject to further optimization. Structural efficiency can be improved by integrating YOLO-World, as discussed in Section \ref{subsubsec:target_localization}, or by increasing the serial communication baud rate from 115200 to a higher operational rate, thereby minimizing the active transmission time between components.

\section{Discussion}
\label{sec:discussion}

The proposed system demonstrates that the integration of ultra-low-power sentinel sensing with high-fidelity local multimodal reasoning can significantly enhance autonomous underwater monitoring capabilities in network-isolated/limited environments. By decoupling continuous environmental surveillance from intensive data processing, the architecture achieves a robust balance between energy efficiency and scientific insight generation. This section discusses the implications of our findings, highlighting the advantages of the local Retrieval-Augmented Generation (RAG) pipeline, while acknowledging operational trade-offs such as inference latency. We also explore how this framework can be extended to diverse marine applications beyond standard biodiversity surveys.

\subsection{Advantages of Local Multimodal RAG and Adaptive Identification}
A key strength of the proposed architecture is its ability to perform complex taxonomic identification without relying on cloud connectivity or pre-trained closed-set classifiers for every species. The use of BioCLIP/OpenCLIP embeddings within a local ChromaDB vector database enables a flexible, retrieval-based approach to species recognition. This method allows researchers to dynamically update reference collections based on specific mission objectives (e.g., focusing on invasive species in one zone and endemic fauna in another) without retraining the underlying neural networks.

The adaptive nature of this identification layer is particularly valuable for exploratory missions where target taxa may vary or remain unknown until deployment. By combining centroid-based similarity search with linear SVMs, the system provides hierarchical taxonomic hypotheses (from class to species level) accompanied by confidence scores and supporting visual evidence. This transparency allows scientists to verify identifications directly at the edge, reducing reliance on post-deployment manual annotation. Furthermore, the integration of textual metadata—such as scientific descriptions and habitat information—into the embedding space enables semantic queries that go beyond simple image matching, fostering a deeper understanding of ecological contexts through local RAG pipelines. Another possibility is using the localization and identification layers to count fish: by cropping and then identifying the species, it will be possible to keep a record and store the count as metadata.

\subsection{Agentic Reasoning: Balancing Intelligence and Latency}
The multi-agent framework (RouterAgent, AnalystAgent, Reporting Agent) introduces a layer of autonomous decision-making that transforms raw sensor data into structured scientific knowledge. While the execution of large language models (LLMs) on edge hardware like the Jetson Orin NX inherently involves higher latency compared to simple inference tasks, this overhead is justified by the value-added services provided: natural-language query handling, automated report generation, and dynamic resource management.

The RouterAgent’s ability to intelligently route queries between visual retrieval, textual analysis, and structured computation ensures that computational resources are allocated efficiently. For instance, deterministic statistical requests are handled via the AnalystAgent using sandboxed Python execution, bypassing the LLM when unnecessary. Although real-time interaction may experience delays due to model loading and generation times, this latency is acceptable for non-critical scientific inspection tasks (Mission Mode) where depth of analysis takes precedence over millisecond-level responsiveness. The system’s design prioritizes comprehensive situational awareness and autonomous reporting capabilities, which are crucial for long-term deployments where human intervention is limited.

\subsection{Energy Efficiency Through Hierarchical Activation}
The hierarchical master--satellite topology effectively mitigates the high energy costs associated with continuous GPU usage. By keeping the Jetson Orin NX powered off except during scheduled cycles or event-triggered activations, the system extends operational duration significantly compared to always-on architectures. The MAX78000/78002 sentinels provide reliable first-stage filtering for visual and acoustic events, ensuring that high-performance processing is reserved for scientifically relevant occurrences.

While the sentinel models operate under strict hardware constraints, their optimized architectures achieve sufficient accuracy for trigger-based activation without excessive false positives. The Energy Management Agent further enhances sustainability by dynamically adjusting system parameters based on remaining battery capacity and mission priorities. Although current energy estimation relies on heuristic approximations, this proactive management strategy demonstrates the feasibility of self-regulating power consumption in remote underwater environments.

\subsection{Limitations of the current framework}
Even though the current system offers many benefits in terms of power consumption, it is still limited by the currently available hardware and software. For instance, the MAX78000 and MAX78002 are highly efficient in terms of power consumption, outperforming almost any available microcontroller in energy per inference. However, they present several issues that may slow down future deployment. First, the hardware accelerator is limited exclusively to CNN models—and specifically, only those that respect the constraints presented in Section \ref{sec:sentinel_subsystems}, making it difficult to deploy models that are already available online. Another problem is the SDK; while most standard chips use classical deployment through ONNX, the Analog Devices chips require training and deployment through a specialized SDK, which slows down the deployment process. On the other hand, currently available VLM (or classical LLM) models under 4B parameters are still not highly reliable, especially for agentic tasks. In the future, more specialized small models may appear, and the current model-agnostic implementation through Ollama will make upgrading and testing new models easier.
\subsection{Scalability and Concrete Underwater Applications}
Beyond standard fish detection and marine mammal acoustic monitoring, the proposed architecture is highly scalable to other critical underwater applications:

\begin{itemize}
    \item \textbf{Invasive Species Monitoring:} The adaptive reference collection feature allows rapid deployment against specific invasive threats (e.g., Lionfish or Green Crab) by simply uploading targeted image/text references. The system can automatically alert researchers when high-confidence detections occur, enabling timely management interventions.

    \item \textbf{Benthic Habitat Assessment:} By integrating visual similarity search with taxonomic metadata, the system can classify and quantify benthic communities (corals, sponges, seagrasses) over time. The RAG pipeline can generate automated health reports based on changes in species composition or coverage area detected by the sentinels.

    \item \textbf{Anthropogenic Activity Tracking:} The multimodal nature of the system supports detection and classification of human-made objects (e.g., fishing gear, debris, or vessels). Acoustic signatures combined with visual confirmation can help monitor illegal fishing activities or assess pollution levels in protected marine areas.

    \item \textbf{ROV/AUV Integration:} The lightweight, modular design is suitable for integration into Remotely Operated Vehicles (ROVs) or Autonomous Underwater Vehicles (AUVs). In this context, the system could provide real-time situational awareness to pilots by highlighting points of interest and generating concise mission summaries upon surfacing.

    \item \textbf{Maritime Surveillance Buoy:} The system can also be deployed on a floating buoy, which guarantees a continuous power source to deliver more frequent and higher-quality reports, process more queries, and utilize hybrid and mission modes more often. Additionally, the energy management agent can be upgraded to account for variable battery charging and discharging cycles, allowing it to optimize the activity of both the sentinel nodes and the master node based on solar patterns.
\end{itemize}

Future work will focus on refining the current pipeline by exploring efficient acoustic embedding techniques—such as CLAP and SAM Audio—for fully multimodal event correlation. Works on seismic detection models were already initiated; however, developing reliable edge-deployable models remains challenging due to training data scarcity and tight microcontroller memory and compute constraints, Consequently, this component has been excluded from the current manuscript pending further validation. Additionally, we aim to optimize the LLM inference pipeline. To eliminate image processing overhead during inference, we will leverage pre-stored CLIP embeddings directly by introducing a projector that bridges the vision and Qwen embedding spaces, while evaluating newly developed models under 4B parameters. Finally, we plan to substitute general biological foundation models with specialized marine domain architectures like AquaticCLIP \cite{alawode2025aquaticclipvisionlanguagefoundationmodel} to extract more granular aquatic environmental features.

In conclusion, this work presents a viable and flexible framework for intelligent underwater monitoring, specifically engineered to operate effectively under severe energy and bandwidth restrictions. By leveraging local multimodal RAG and adaptive identification, the system empowers researchers to obtain actionable ecological insights from remote deployments with minimal energy consumption and communication overhead. Its modular design opens new possibilities for diverse marine observation tasks, bridging the gap between raw data collection and automated scientific interpretation at the edge.

\section{Acknowledgments}
\label{sec:acknowledgments}

This work was conducted as part of an internship funded by the French Research Institute for Exploitation of the Sea (Ifremer). The authors gratefully acknowledge Ifremer for its financial support and computational resources, specifically the Datarmor HPC cluster that facilitated model training. Additionally, we express our gratitude to the creators of the WildFish dataset for granting access to their dataset, and to Mathieu Léonardon for endorsing the preprint.

\printcredits

\bibliographystyle{cas-model2-names}
\bibliography{cas-refs}

\end{document}